\documentclass[12pt]{article}

\usepackage{graphicx} 
\usepackage{dcolumn}
\usepackage{bm}
\usepackage{tabularx}

\def\iotabar{\lower3pt\hbox{$\mathchar'26$}\mkern-8mu\iota}

\begin{document}
\bibliographystyle{unsrt}

\title{The role of magnetic islands in modifying long range temporal correlations of density fluctuations and local heat transport}

\author{ \small{B.Ph.~van Milligen$^1$, T.~Estrada$^1$, L.~Garc\'ia$^2$, D.~L\'opez~Bruna$^1$,} \\
\small{B.A.~Carreras$^3$, Y.~Xu$^4$, M.~Ochando$^1$, C.~Hidalgo$^1$,} \\
\small{J.M.~Reynolds-Barredo$^2$, A. L\'opez Fraguas$^1$}\\
\small{$^1$ CIEMAT - Laboratorio Nacional de Fusi{\'o}n, Avda.~Complutense 40, 28040 Madrid, Spain}\\
\small{$^2$ Universidad Carlos III, 28911 Legan\'es, Madrid, Spain}\\
\small{$^3$ BACV Solutions, 110 Mohawk Road, Oak Ridge, Tennessee 37830, USA}\\
\small{$^4$ Southwestern Institute of Physics, P.O. Box 432, Chengdu 610041, People's Republic of China}}

\maketitle

\begin{abstract}
This work explores the relation between magnetic islands, long range temporal correlations and heat transport.
A low order rational surface ($\iotabar = 3/2$) was purposely scanned outward through an Electron Cyclotron Resonance Heated (ECRH) plasma in the TJ-II stellarator.
Density turbulence and the poloidal flow velocity (or radial electric field) were characterized using a two channel Doppler Reflectometer. 
Simultaneously, the ECRH power was modulated to characterize heat transport, using measurements from a 12 channel Electron Cyclotron Emission diagnostic.
A systematic variation of the poloidal velocity was found to be associated with the stationary $\iotabar = 3/2$ magnetic island.
Inside from the rational surface, the Hurst coefficient, quantifying the nature of long-range correlations, was found to be significantly enhanced.
Simultaneously, heat transport was enhanced as well, establishing a clear link between density fluctuations and anomalous heat transport. 
The variation of the Hurst coefficient was consistent with a magnetohydrodynamic turbulence simulation.
\end{abstract}
\vspace{3em}

\section{Introduction}

In this work, we study the relation between rational surfaces, long range temporal correlations and heat transport in the low shear stellarator TJ-II.
The relationship between rational surfaces and heat transport has been explored in a number of earlier works, both at TJ-II~\cite{Hidalgo:2001b,Castejon:2004} and other stellarator devices \cite{Ida:2004}.
In tokamaks, low order rational surfaces are often associated with Internal Transport Barriers or the reduction of (heat) transport~\cite{Connor:2004,Austin:2006}.
However, very little is known about the relation between rational surfaces and the Hurst coefficient~\cite{Pan:2015}, and this work will focus specifically on this issue.
The Hurst coefficient is a measure of the degree of long range temporal correlations~\cite{Mandelbrot:1969,Carreras:1998b,Sanchez:2006}, and hence an important quantifier of anomalous transport. 

In the experiments presented here, the position of the rational surfaces is slowly scanned outwards by means of the external induction of Ohmic current~\cite{LopezBruna:2009}, allowing the exploration of various quantities as a function of the magnetic flux by diagnostics observing at a fixed position.
From Doppler reflectometer (DR) measurements, it is found that a rational surface belonging to a low order rational (3/2) is associated with a localized variation of the radial electric field.
When this rational surface crosses the DR measurement location, the Hurst coefficient obtained from the DR density fluctuation amplitude is modified significantly.

In the same discharges, we also characterize heat transport using the modulation of centrally deposited Electron Cyclotron Resonance Heating (ECRH) combined with radially distributed Electron Cyclotron Emission (ECE) temperature measurements.
We observe that transport is locally enhanced when the DR Hurst coefficient is increased. 
Eight similar discharges are analyzed, showing that the obtained results are systematic and reproducible.

The reported observations establish a clear link between 
(a) the modification of density fluctuation properties near a rational surface in the gradient region, associated with long range memory effects, and 
(b) enhanced heat transport.

The interpretation of these observations finds support from a model~\cite{Garcia:2001}, showing that the radial electric field structure and the Hurst coefficient are both modified close to the rational surface.

This paper is structured as follows. 
In Section \ref{techniques}, the experimental setup is discussed.
In Section \ref{experiments}, we present the experimental results.
In Section \ref{modeling}, we present the modeling results and interpretation.
In Section \ref{discussion}, we discuss the results, and in Section \ref{conclusions}, we draw some conclusions.

\clearpage
\section{Experimental set-up and techniques used}\label{techniques}

In this work, we study discharges heated by Electron Cyclotron Resonant Heating (ECRH).
In these discharges, the plasma has a relatively low line average density of $\overline n_e \simeq 0.5 \cdot 10^{19}$ m$^{-3}$, so that it is in the electron root state, slightly below the critical density of the electron to ion root confinement transition at TJ-II~\cite{Hidalgo:2006b,Velasco:2012}.

\subsection{Modulated ECRH and ECE measurements}\label{ecrh}

The ECRH system consists of two gyrotrons with a frequency of 53.2 GHz, allowing the injection of up to $2 \times 300$ kW of heating power~\cite{Fernandez:2009}.
In the ECRH experiments, both ECRH systems were launching into the core of the plasma; the half width of the deposition profile being $w_{\rm ECRH} \simeq 3$ cm~\cite{Eguilior:2003}. 
One was providing continuous injection of approximately 250 kW, whereas the second system was modulated at low power ($\sim 50$ kW), at a modulation frequency of 360 Hz and with a duty cycle of 30 \%, in order to allow quantifying heat transport.

TJ-II disposes of a 12 channel Electron Cyclotron Emission (ECE) detection system to measure the local electron temperature $T_e$ at up to 12 different radial positions along the midplane on the high magnetic field side of the plasma (at $\phi = 315^\circ$), covering a significant part of the plasma minor radius, with a radial resolution of about 1 cm~\cite{delaLuna:2001}. 
The local modulation amplitude and phase (relative to a central ECE channel) are determined using a Fourier analysis of the $T_e(r,t)$ data.

Assuming transport is diffusive it is possible, in principle, to extract diffusive transport parameters (such as the effective diffusivity and convective velocity) from these data.
However, we refrain from doing so, as the uncertainties introduced by assuming a specific transport model, as well as the ill-posed nature of the problem, do not allow obtaining a clear and unambiguous interpretation of the transport processes inside the plasma from such an inversion~\cite{Lopes:1995,Sattin:2012}.
Instead, we will represent the amplitude and relative phase directly, noting that the gradient of the relative phase is directly linked to the speed of the transport~\cite{Jacchia:1991,Mantica:2006b}.

\subsection{Doppler reflectometry}

The two-channel Doppler reflectometer of TJ-II~\cite{Happel:2009} is located in a top viewport at $\phi \simeq 337^\circ$.
In Doppler reflectometry, a finite tilt angle is purposely introduced between the incident probing beam and the normal to the reflecting cut-off layer, and the Bragg back-scattered signal is measured~\cite{Hirsch:2001}.
The amplitude of the recorded signal is a measure of the intensity of the density fluctuations, $\tilde n$. 
As the plasma rotates in the reflecting plane (a flux surface), the scattered signal experiences a Doppler shift. 
The size of this shift is directly proportional to the rotation velocity of the plasma turbulence perpendicular to the magnetic field lines, $v_\perp$, and therefore to the plasma background $E \times B$ velocity, provided the latter dominates over the phase velocity of density fluctuations~\cite{Estrada:2009}.
The Doppler reflectometer signals, sampled at 10 MHz, allow determining $\tilde n$ and $v_\perp$ with high temporal and spatial resolution at two radial positions~\cite{Estrada:2012}.
In the discharges analysed here, channel 1 is characterized by: $\rho \simeq 0.75$, $k_\perp \simeq 6$ cm$^{-1}$ and channel 2 by: $\rho \simeq 0.68$, $k_\perp \simeq 6.7$ cm$^{-1}$.

\subsection{Sweep and tracking of rational surfaces}\label{tracking}

\begin{figure}\centering
  \includegraphics[trim=0 0 0 0,clip=,width=12cm]{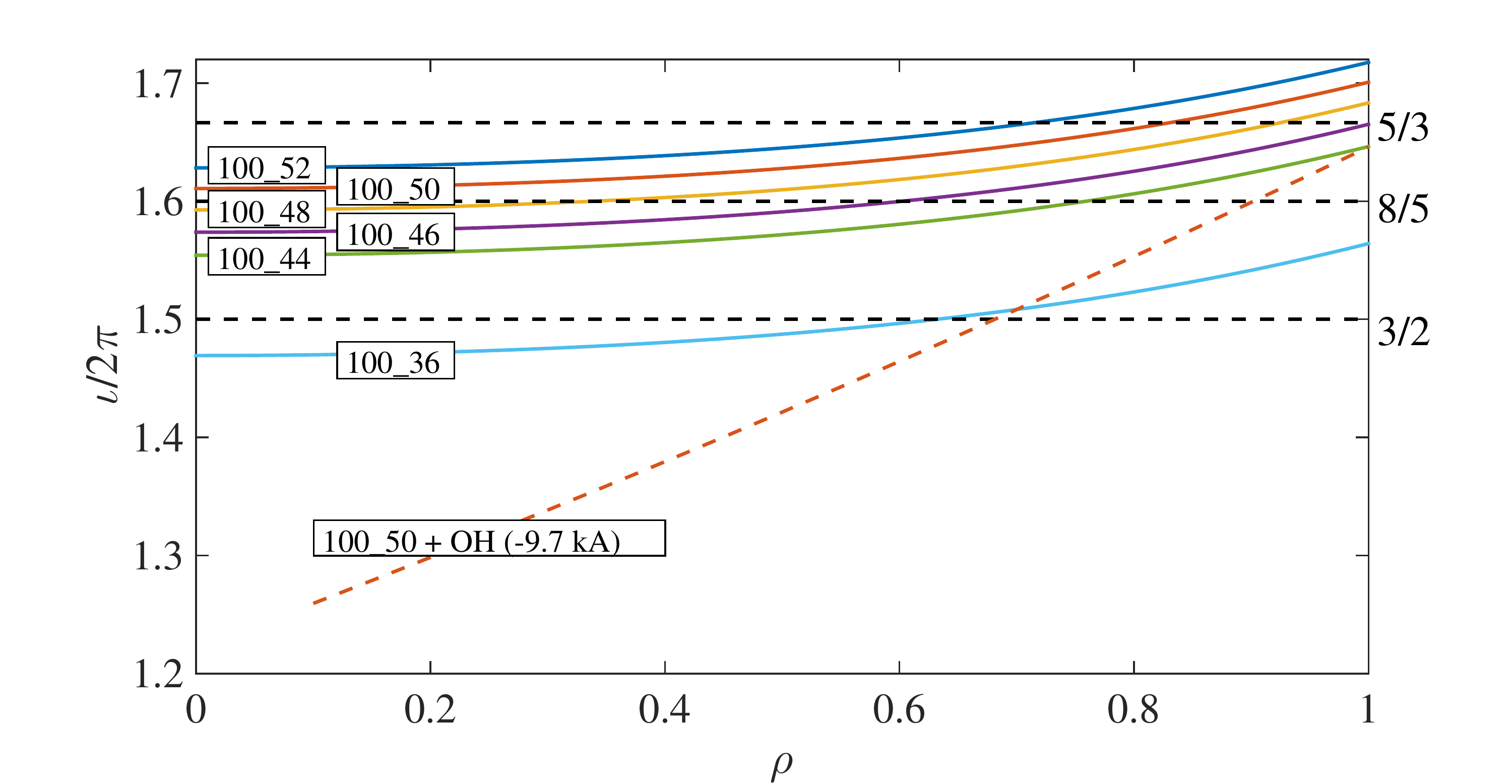}
\caption{\label{iota}Profiles of the rotational transform for several magnetic configurations and an example of the modification produced by an inducted Ohmic current.}
\end{figure}

In these ECRH experiments, a number of magnetic configurations were used with similar (low) magnetic shear, but a slightly different value of rotational transform, $\iotabar$. 
By inducing an Ohmic current in the plasma, the rotational transform profile is modified, as shown in Fig.~\ref{iota}.
The induced Ohmic current moves the rational surfaces (such as $\iotabar = 3/2$) outward.
Simultaneously, the magnetic shear is enhanced, although it remains rather low.

\begin{figure}\centering
  \includegraphics[trim=0 0 0 0,clip=,width=11cm]{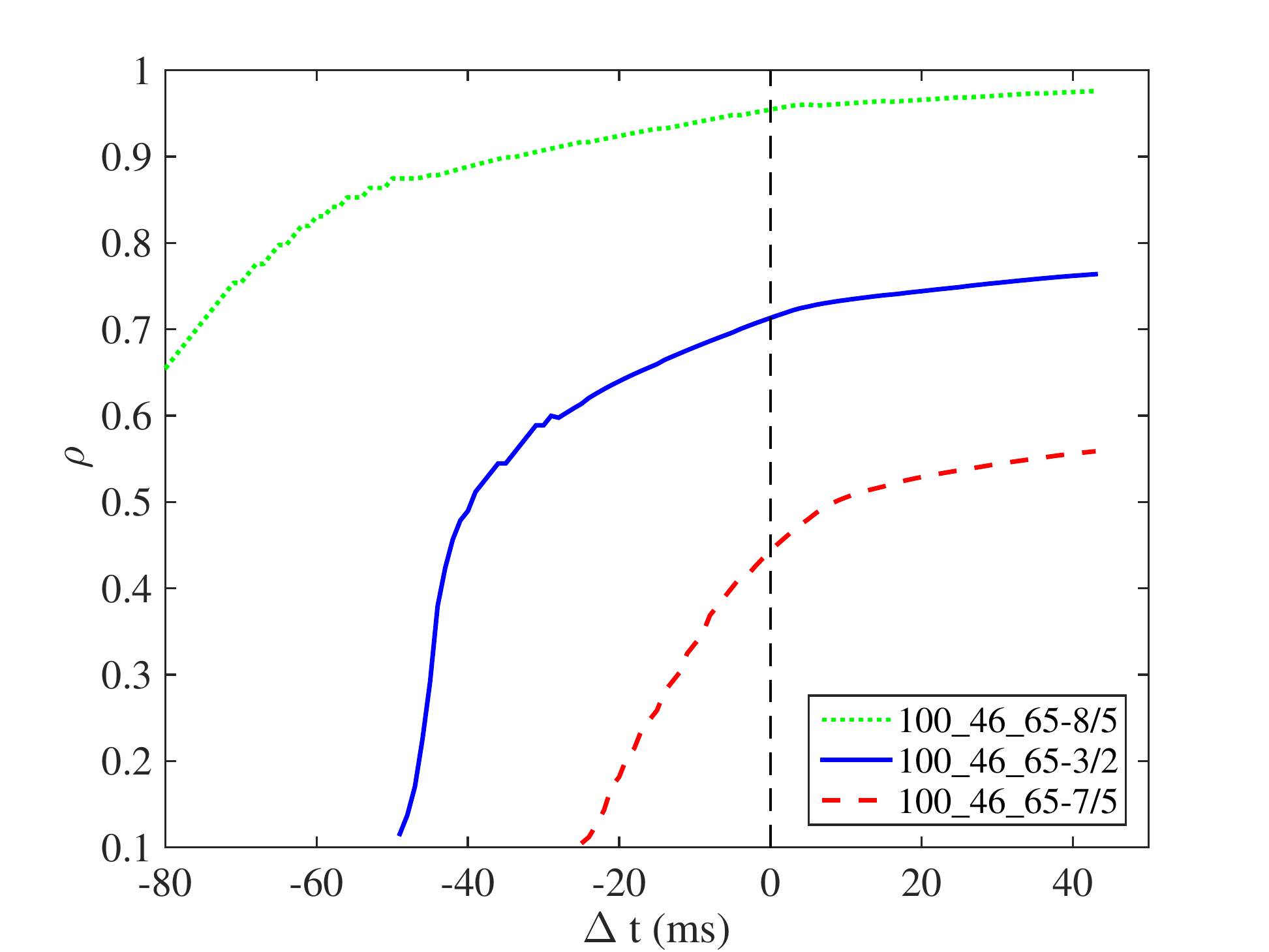}
\caption{\label{track}Temporal evolution of the radial location of three rational surfaces during the scan of the plasma current ($\iotabar = 8/5, 3/2$, and $7/5$).}
\end{figure}

For the interpretation of experimental data obtained in such discharges, a reconstruction of the actual rotational transform profile (and the location of the rational surfaces) during the sweep of the Ohmic current is useful.
This is obtained by iteratively updating equilibria calculated by VMEC~\cite{Hirshman:1986}, using an estimate for the Spitzer resistivity based on experimental data~\cite{Strand:2001}.
Details of this calculation will be reported in a separate publication.
An example result is shown in Fig.~\ref{track}, with respect to a reference time $\Delta t = 0$ to be defined in the following. 
However, this evolution will depend on the precise shape of the plasma current profile, which is not measured directly, leading to some uncertainty in the calculation. 
The interpretation of the data reported in the following will therefore not rely on this calculation in detail, although it serves as a guide to what one may expect to find in the experimental situation.

\begin{figure}\centering
  \includegraphics[trim=0 0 0 0,clip=,width=16cm]{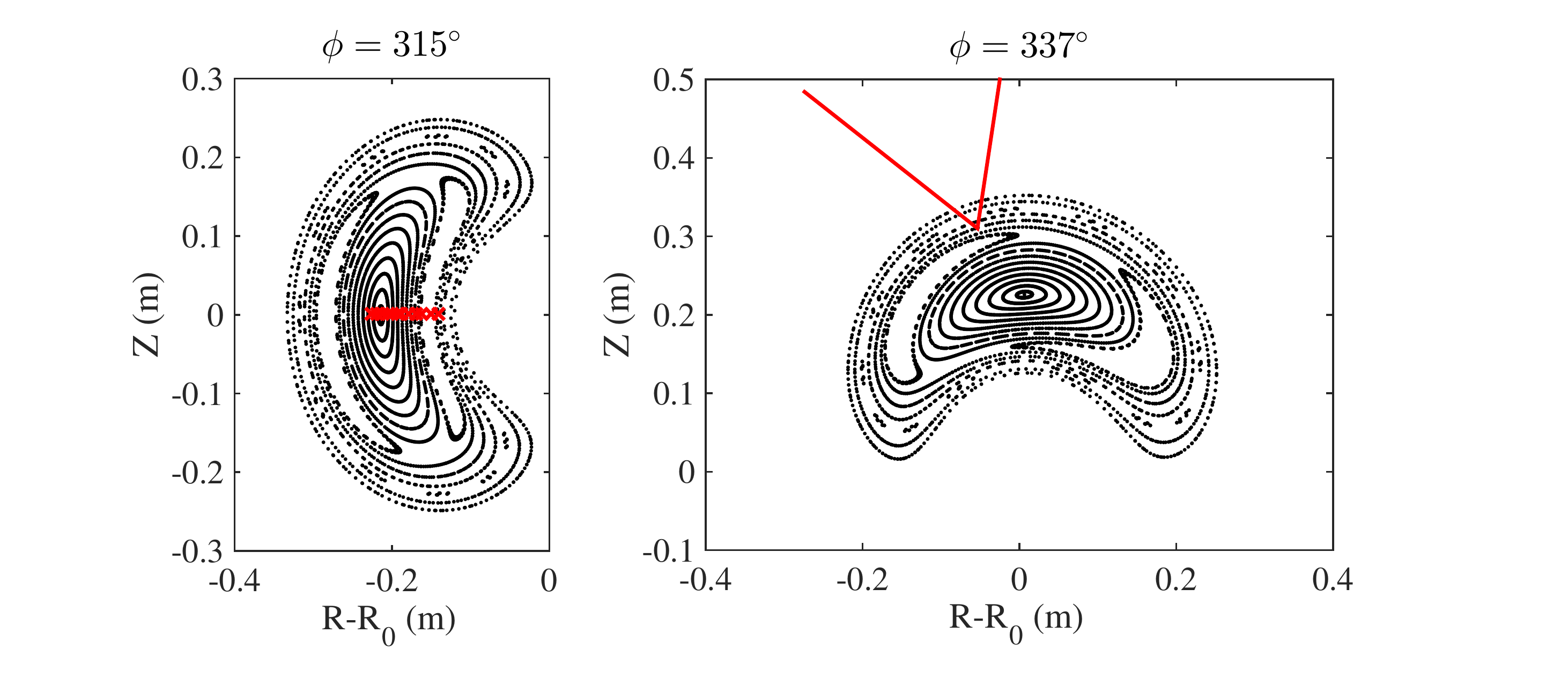}
\caption{\label{poincare}Poincar\'e plots for the vacuum configuration 100\_36 (cf.~Fig.~\ref{iota}), in $\phi = $ constant planes, indicating the positions of the ECE diagnostic (left) and the DR diagnostic (right). $R_0 = 1.5$ m.}
\end{figure}

In this work, we will focus mainly on the $\iotabar = 3/2$ rational and its associated island.
Fig.~\ref{poincare} shows Poincar\'e plots for the poloidal cross sections where the ECE diagnostic and the DR diagnostic are located.
The $\iotabar = 3/2$ island has been generated by adding an appropriate resonant perturbation to the vacuum magnetic field corresponding to the magnetic configuration labelled 100\_36 in Fig.~\ref{iota}.
Neither the configuration, the width of the island, nor its precise poloidal location need coincide closely with the actual (dynamically varying) experimental situation at any point in time; however, we show the figure as a guide to the interpretation of the measurements reported in the following section.

\subsection{The Hurst coefficient}\label{hurst}

In this work, we will be using the Hurst coefficient as a measure of long-range correlations.
This coefficient provides an estimate of the degree to which transport deviates from normal diffusive transport.
We recall that the Hurst coefficient lies in the range $0 \le H \le 1$. 
When $H=0.5$, one has ordinary diffusive transport; when $0 \le H < 0.5$, one has antipersistence of the fluctuations, associated with subdiffusive transport, and when $0.5 < H \le 1$, persistence of the fluctuations, associated with superdiffusive transport~\cite{Carreras:1998b,Sanchez:2006}.

The Hurst coefficient is calculated from the density fluctuation amplitude, $|\tilde n|$, measured using the DR, using the method described in \cite{Carreras:1999b}. 
We calculate the rescaled range graph of $R/S$ using overlapping time windows with a length of 10 ms, and evaluate $H$ from the logarithmic slope of $R/S(\tau)$ over the time lag range of $0 \le \log_{10} \tau \le 1$ ($\tau$ in ms). 
The value of $H$ thus obtained corresponds to asymptotic times, close to the confinement time of the plasma.
Note that the DR sampling rate is 10 MHz, so that 10 ms of data corresponds to $10^5$ data points, providing ample statistics for the determination of the Hurst coefficient.
Even though the plasma state is varying constantly due to the ramped plasma current, these changes are sufficiently slow to consider the density fluctuation amplitude measurement to be essentially in steady state during the analysis time windows of 10 ms.
Modifying the length of these time windows by a factor of two hardly affects the results, confirming this statement.
Furthermore, MHD mode activity is weak and is found not to affect the $R/S$ statistic.

\clearpage
\section{Experiments}\label{experiments}

\subsection{Doppler reflectometry results}

\begin{figure}\centering
  \includegraphics[trim=0 0 0 0,clip=,width=9cm]{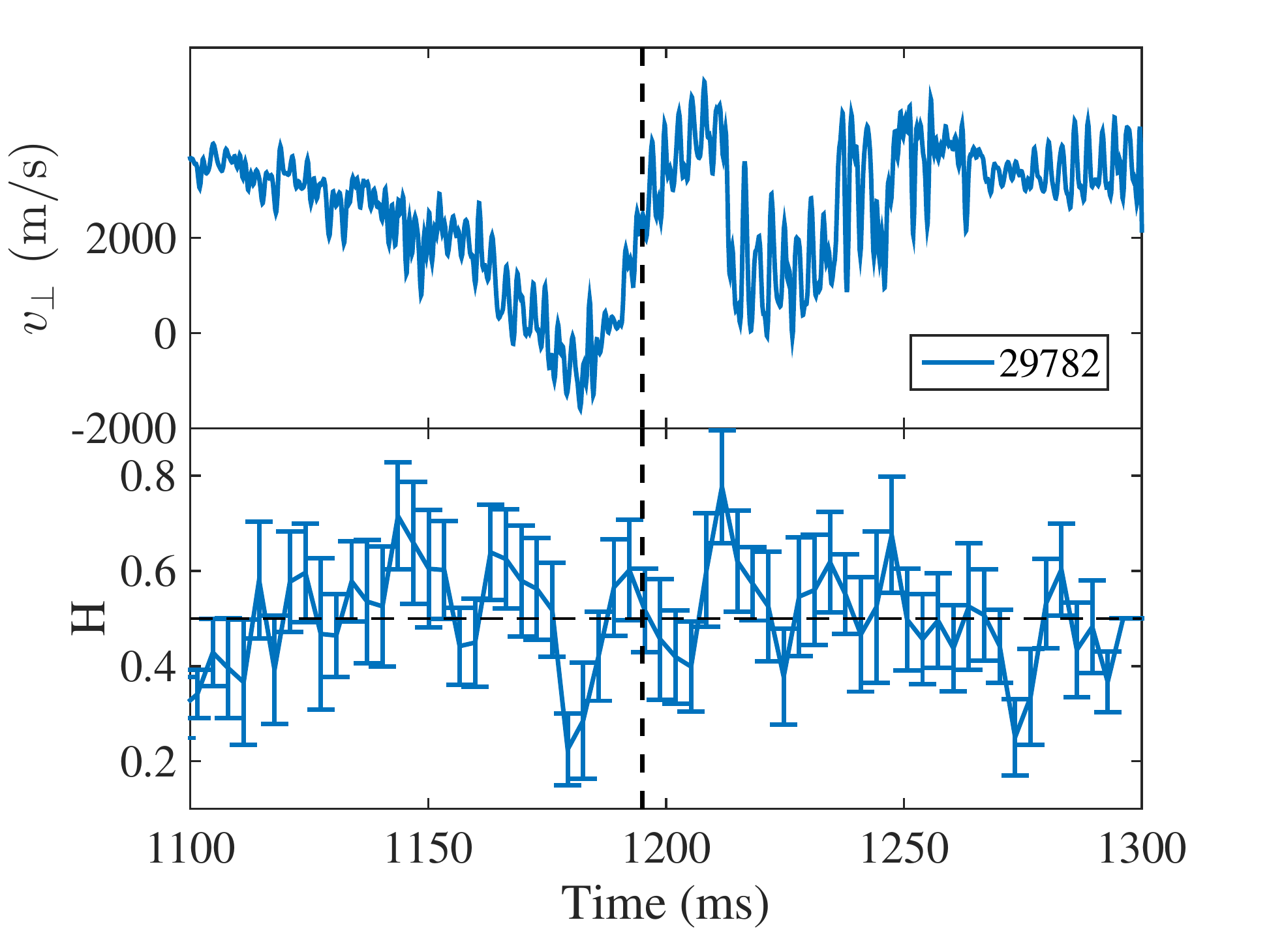}\\
  \includegraphics[trim=0 0 0 0,clip=,width=9cm]{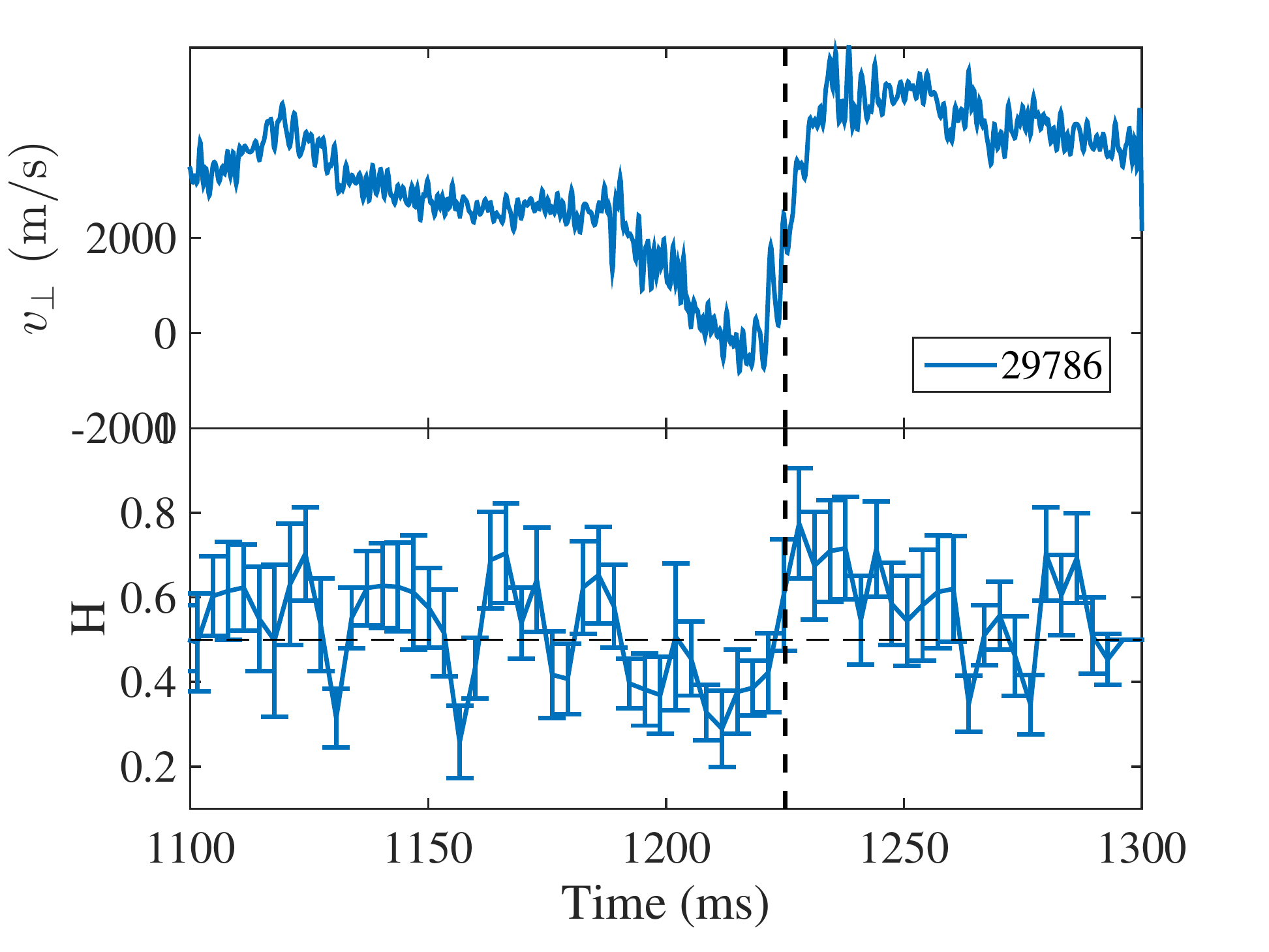}\\
  \includegraphics[trim=0 0 0 0,clip=,width=9cm]{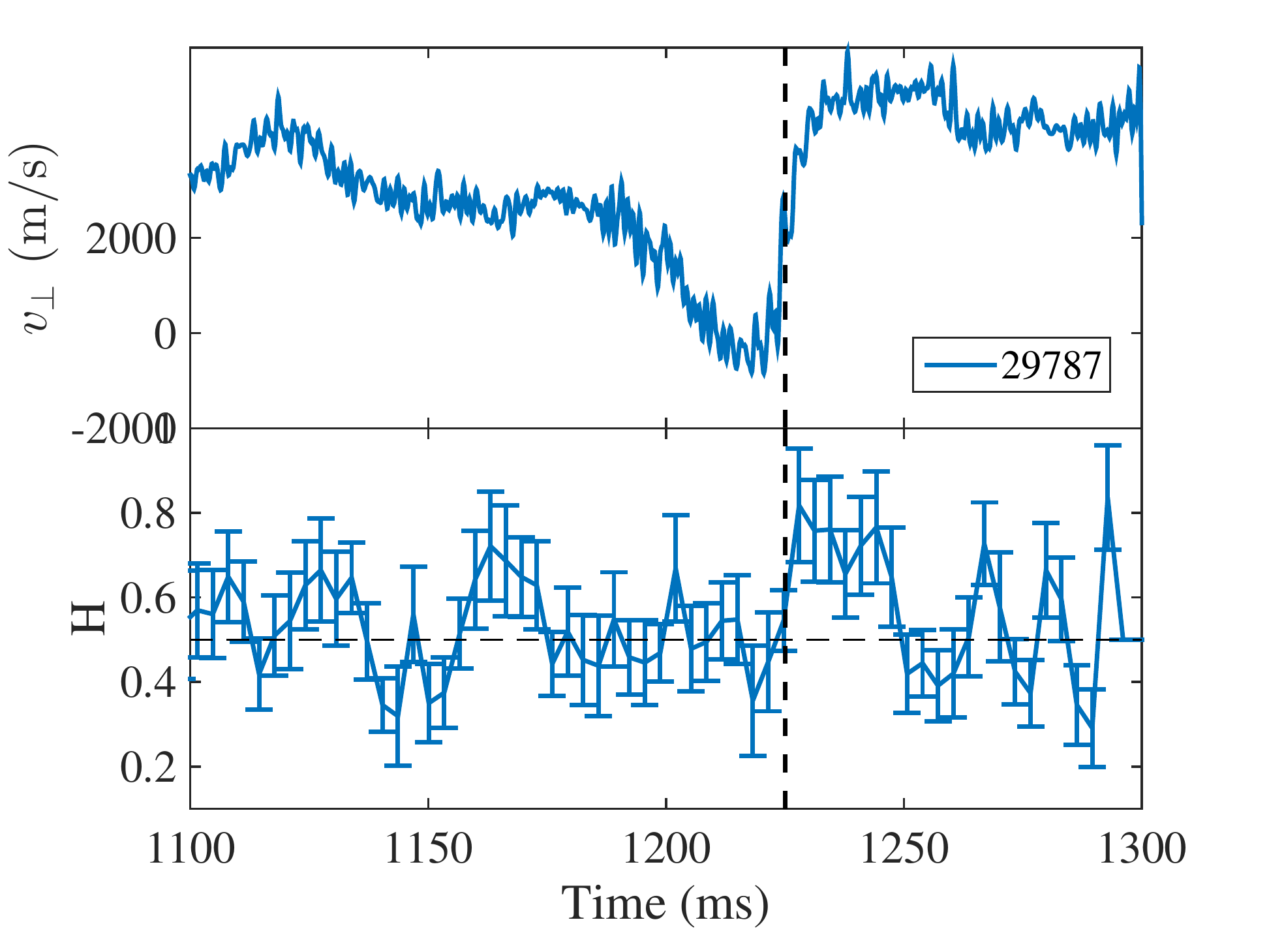}
\caption{\label{COG_Hurst}Smoothed $v_\perp$ signal  and Hurst coefficient calculated from $|\tilde n|$ (three discharges, DR channel 1).}
\end{figure}

Figs.~\ref{COG_Hurst} show examples of the analysis of Doppler reflectometry signals.
Here, $v_\perp$ is computed from the raw complex signal using 64-point bins~\cite{Estrada:2012}.
The figures show $v_\perp$ itself (smoothed to remove noise), showing the mean perpendicular rotation velocity at the observation point, and the Hurst coefficient, $H$, calculated as described in Section~\ref{hurst}.
The 360 Hz modulation seen in $v_\perp$ is a consequence of the modulated ECRH power.

The Hurst coefficient tends to be slightly above $H =0.5$ on average, although during specific time intervals, the value is lower or higher, which may be related to the passage of rationals by the observation point.
In Fig.~\ref{COG_Hurst}, $H$ is particularly low for discharge 29782, $1175 < t < 1185$ ms, coincident with $v_\perp \simeq 0$.
Similarly, for discharge 29786, $H$ is low for $1205 < t < 1220$ ms, again coincident with $v_\perp \simeq 0$.
The dip in $H$ is somewhat less clear for discharge 29787.
As will be clarified in Section \ref{modeling}, the periods of $H < 0.5$ are associated with subdiffusion occurring in `trapping zones', characteristic of magnetic island O-points associated with a low order rational surface.
On the other hand, in Fig.~\ref{COG_Hurst}, $H$ is high for $1210 < t < 1220$ ms in discharge 29782, and for $1225 < t < 1250$ ms in discharges 29786 and 29787.

\begin{figure}\centering
  \includegraphics[trim=0 0 0 0,clip=,width=12cm]{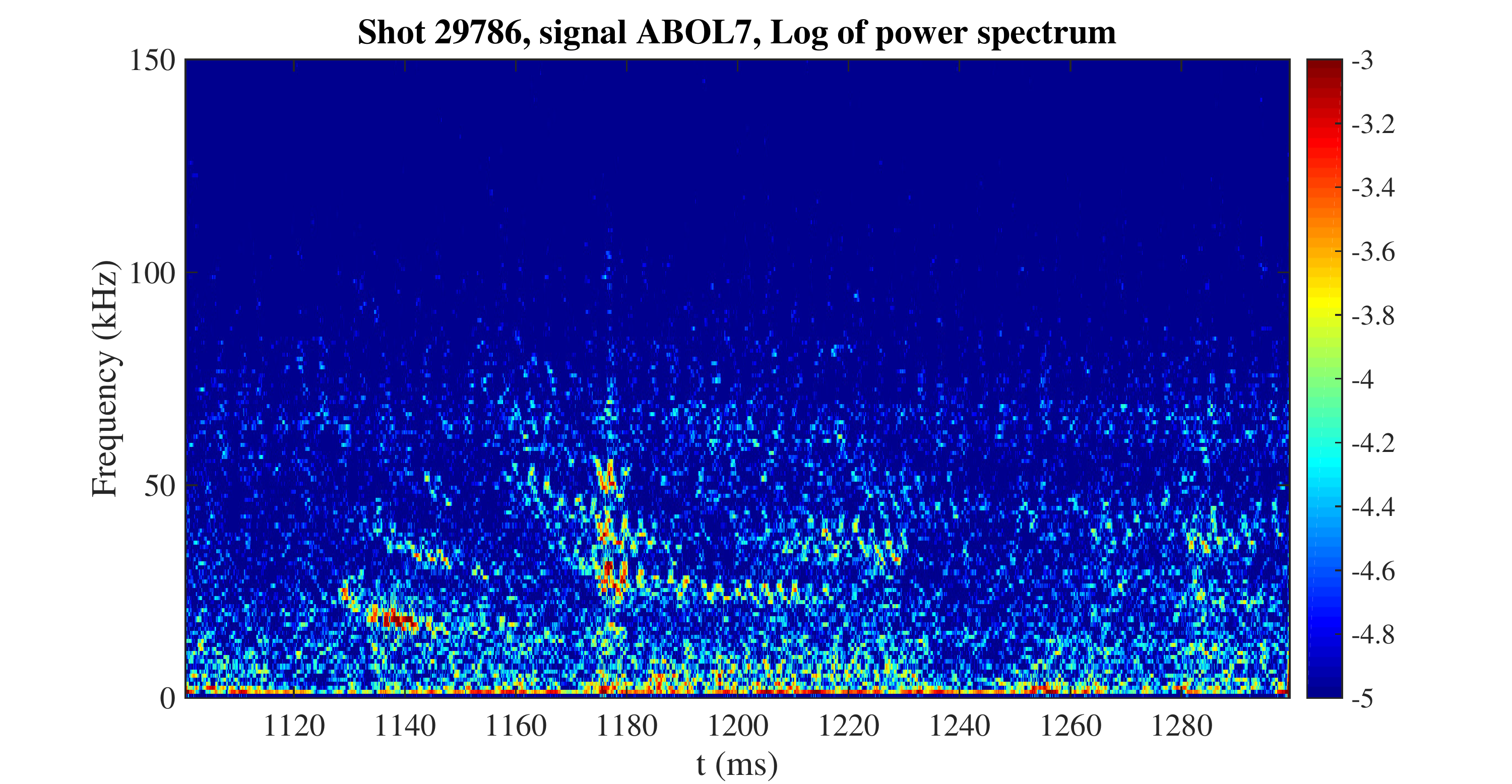}
\caption{\label{29786_ABOL7_spec}ABOL7 spectrum. The overall slowdown of the frequency is due to the rising OH current, slowly moving the rational surfaces outward. The modulation of the mode frequency (at 360 Hz) is due to ECRH modulation (discharge 29786).}
\end{figure}

Fig.~\ref{29786_ABOL7_spec} shows the spectrum of a bolometry channel (ABOL7) for discharge 29786, also shown in Fig.~\ref{COG_Hurst}~\cite{Ochando:2013}.
One observes several modes, including a mode with a frequency of $\sim 30$ kHz, visible in the time interval $1170 < t < 1220$.
The mode frequency is modulated by the ECRH modulation frequency (360 Hz).
The mean frequency of the mode sweeps down slowly because the rational surface is moving outward (to parts of the plasma that are rotating more slowly)~\cite{Milligen:2012}.
The apparent disappearance of this mode at $t \simeq 1220$ is significant, as we will discuss in the following.

To get a clearer view of the temporal evolution of the various quantities, we will use 8 similar discharges to compute averages. 
These discharges are listed in Table~\ref{dischargelist}.
The starting configurations are different, having vacuum $\iotabar$ profiles as shown in Fig.~\ref{iota}, meaning (mainly) that the time of occurrence of the passage of the rational surface past the Doppler reflectometry observation point is different for each discharge. 
Also, the sweep rates of the OH current vary slightly.
To be able to average over such discharges, we must define a reference time, which we take to be the time of steepest slope of $v_\perp$, i.e., the times indicated by the vertical dashed lines in Fig.~\ref{COG_Hurst}.
The reference times are different for each channel, as channel 1 (further outward) responds some 15 ms after channel 2. 

\begin{table}[htdp]
\caption{Discharges and DR reference times}
\begin{center}
\begin{tabular}{|c|c|c|c|}
\hline
Discharge number & Starting configuration &Reference time &Reference time \\
& & (Channel 1)  & (Channel 2) \\
\hline
    29771&100\_44\_64&1160 & 1150\\
    29780&100\_46\_65&1170 & 1160\\
    29782&100\_48\_65&1195 & 1185\\
    29783&100\_48\_65&1185 & 1175\\
    29784&100\_50\_56&1210 & 1190\\
    29785&100\_50\_56&1200 & 1180\\
    29786&100\_52\_66&1225 & 1200\\
    29787&100\_52\_66&1225 & 1205\\
\hline
\end{tabular}
\end{center}
\label{dischargelist}
\end{table}%

\begin{figure}\centering
  \includegraphics[trim=0 0 0 0,clip=,width=12cm]{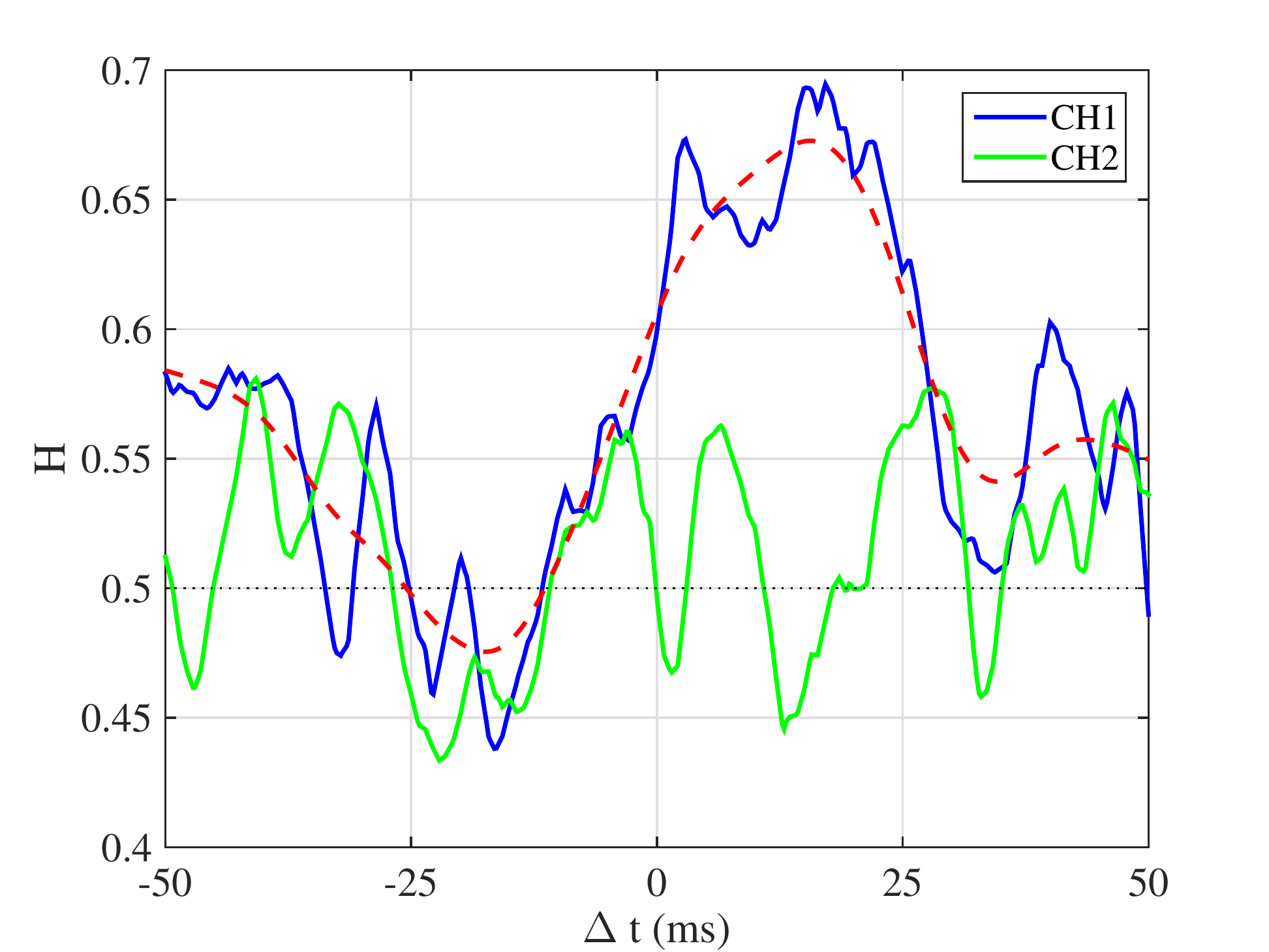}
\caption{\label{mean_COG_H}
The mean Hurst coefficient (calculated from $|\tilde n|$ for 8 discharges) for Doppler reflectometry channels 1 and 2.
The red dashed line indicates the smoothed trend.}
\end{figure}

Fig.~\ref{mean_COG_H} shows the mean evolution of $H$ in a time window centered about the reference time.
The mean Hurst coefficient of channel 1 is close to the value for uncorrelated random noise (diffusion), $H\simeq 0.5$, for $\Delta t < 0$, and attains a peak of about $H=0.65-0.7$ lasting about 25 ms after the steepest descent point of $v_\perp$.
The dip in $H$, observed in Fig.~\ref{COG_Hurst}, is not seen clearly here, due to the mentioned variability of the discharge conditions; but the rise in $H$ seen at $\Delta t > 0$ appears to be quite robust to the averaging procedure.
Channel 2 (further inward) does not show a clear temporal evolution of $H$.

\begin{figure}\centering
  \includegraphics[trim=0 0 0 0,clip=,width=11cm]{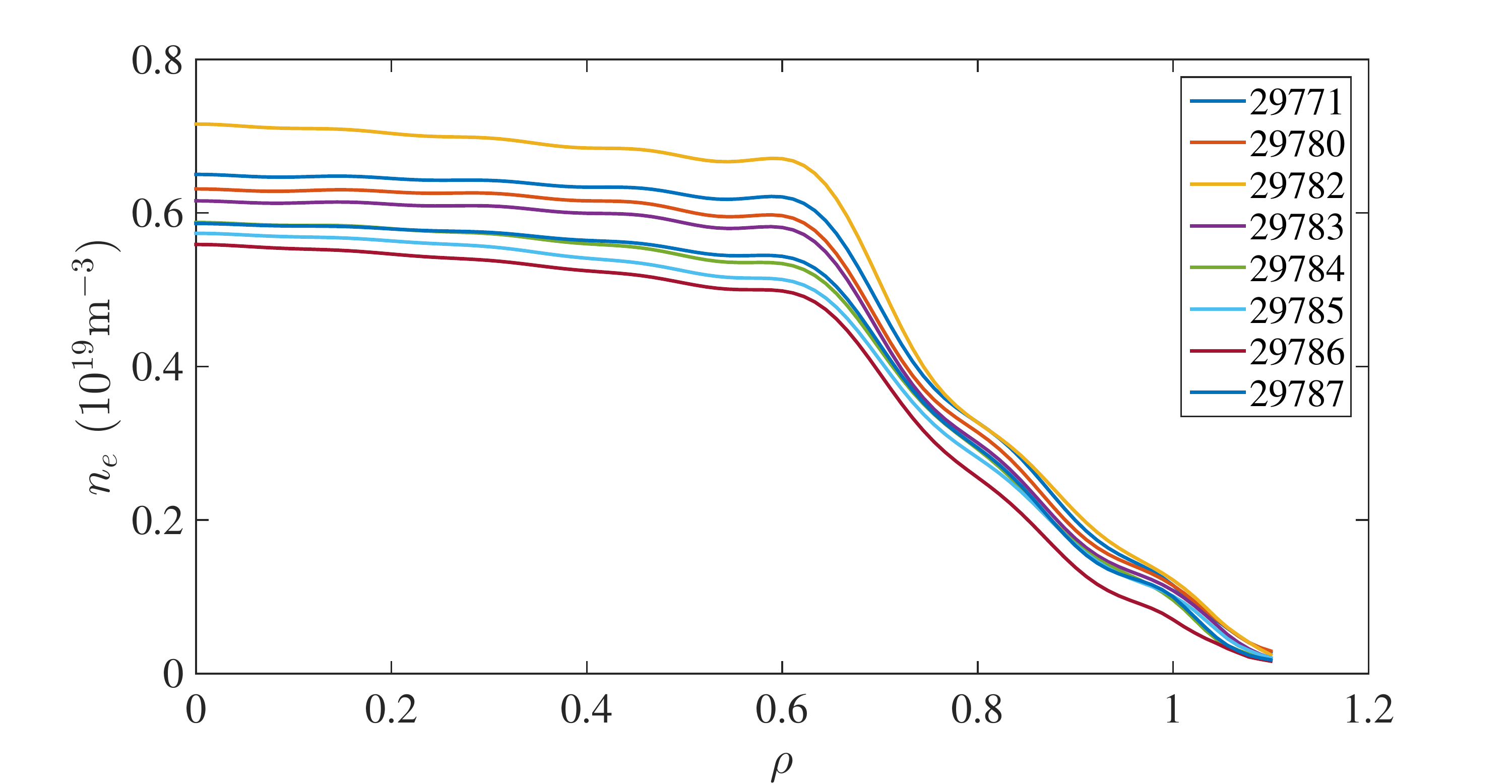}
\caption{\label{mean_ne}Mean electron density profiles for the eight discharges analyzed in this work. The profiles are averaged over $-100 \le \Delta t \le 100$ ms; in this time window, temporal variations are only slight.}
\end{figure}

To understand this situation, Fig.~\ref{mean_ne} shows the mean electron density profiles reconstructed from AM reflectometer~\cite{Estrada:2001} and interferometer data, using a Bayesian technique described elsewhere~\cite{Milligen:2011b}.
The density profiles evolve only little in the relevant time interval.
Clearly, when the rational $\iotabar = 3/2$ is located at a position $\rho < 0.65$, the density gradient at the rational surface is low, whereas it is much higher for $\rho > 0.65$.
As island growth is related to local gradients, we surmise that the corresponding island is initially small, but grows when the rational surface reaches the gradient region, $\rho > 0.65$.
The rational surface is associated both with a dip in $H$ and $v_\perp \simeq 0$ (cf.~Fig.~\ref{COG_Hurst}), the latter being consistent with the stationarity of the island in the density gradient region.
The stationarity of the island for $\rho > 0.65$ may explain the disappearance of mode oscillations observed by bolometry.
After passage of the rational, $H$ clearly increases for channel 1 (well inside the gradient region at $\rho \simeq 0.75$), while it remains modest at channel 2 (at $\rho \simeq 0.68$).

\clearpage
\subsection{ECE}

\begin{figure}\centering
  \includegraphics[trim=0 0 0 0,clip=,width=12cm]{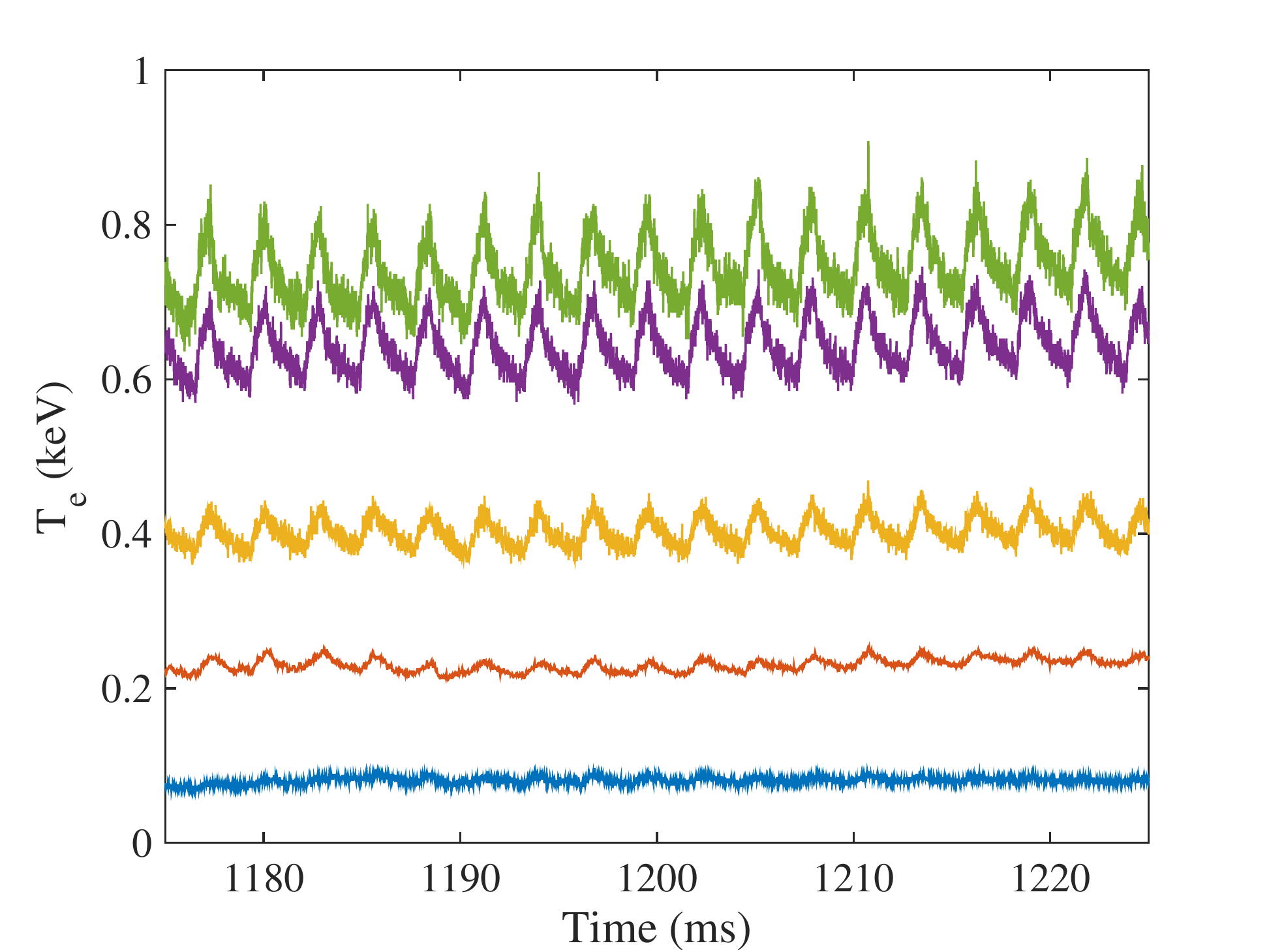}
\caption{\label{29786_Te}ECE data $T_e(\rho,t)$ obtained in an ECRH modulation experiment (discharge 29786, ECE channels $1,3,6,8,11$).}
\end{figure}

\begin{figure}\centering
  \includegraphics[trim=0 0 0 0,clip=,width=12cm]{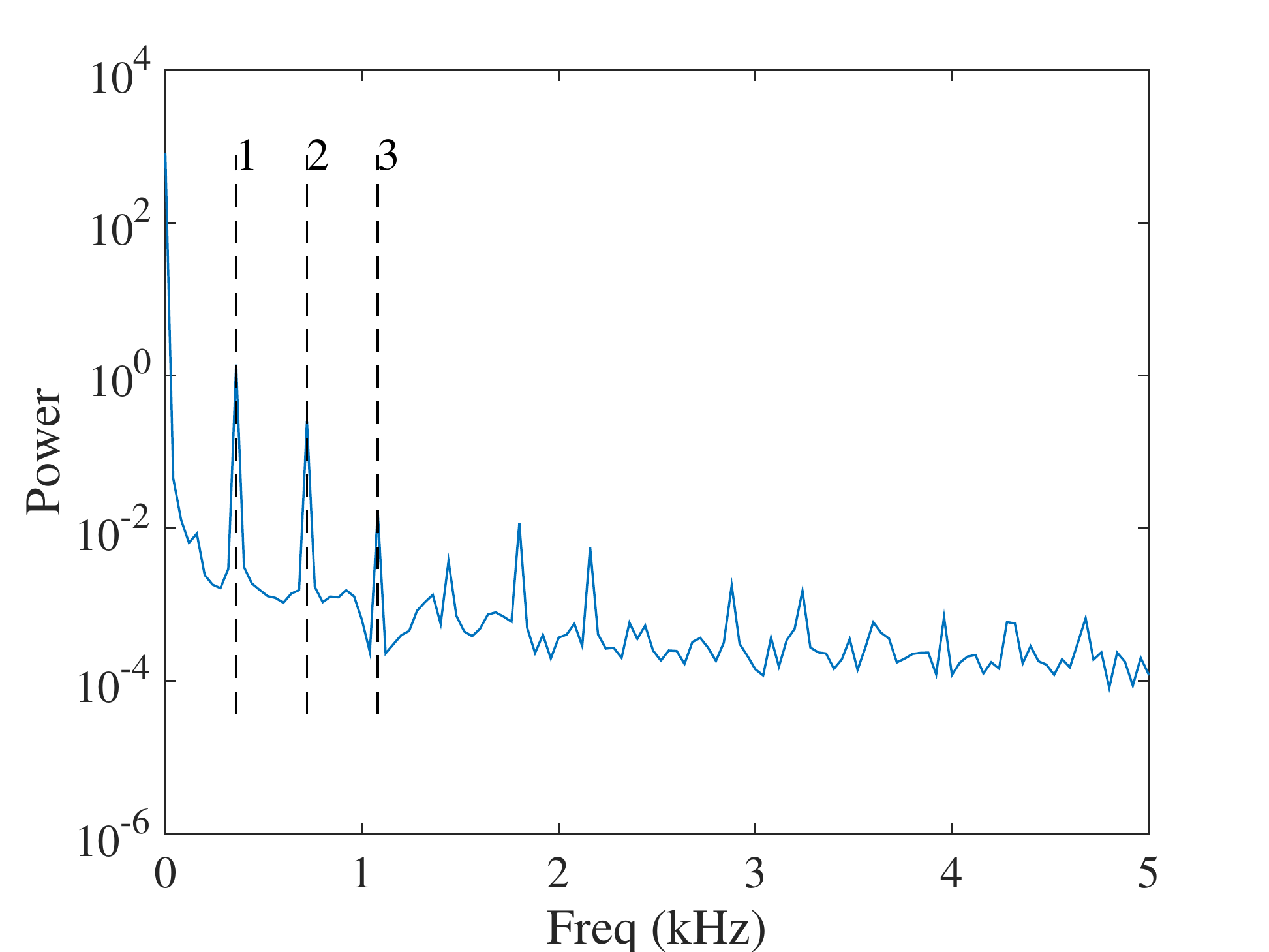}
\caption{\label{29786_Te_spec}Power spectrum of the $T_e(\rho,t)$ data shown in Fig.~\ref{29786_Te}, averaged over radius.}
\end{figure}

\begin{figure}\centering
  \includegraphics[trim=0 0 0 0,clip=,width=12cm]{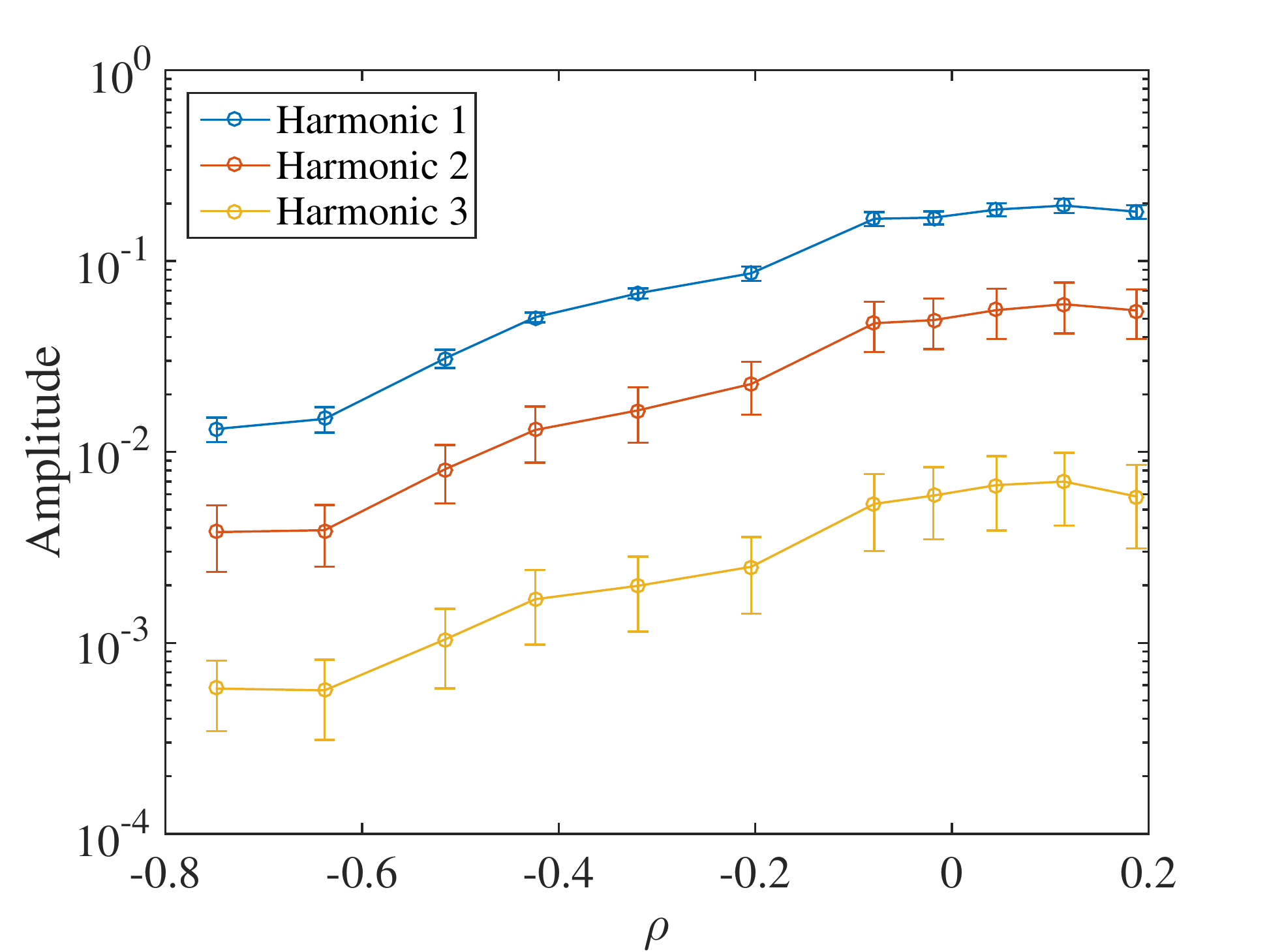}
\caption{\label{29786_amp}Amplitude, $A$, of the first three harmonics versus radius, calculated from the $T_e(\rho,t)$ data shown in Fig.~\ref{29786_Te}.}
\end{figure}

\begin{figure}\centering
  \includegraphics[trim=0 0 0 0,clip=,width=12cm]{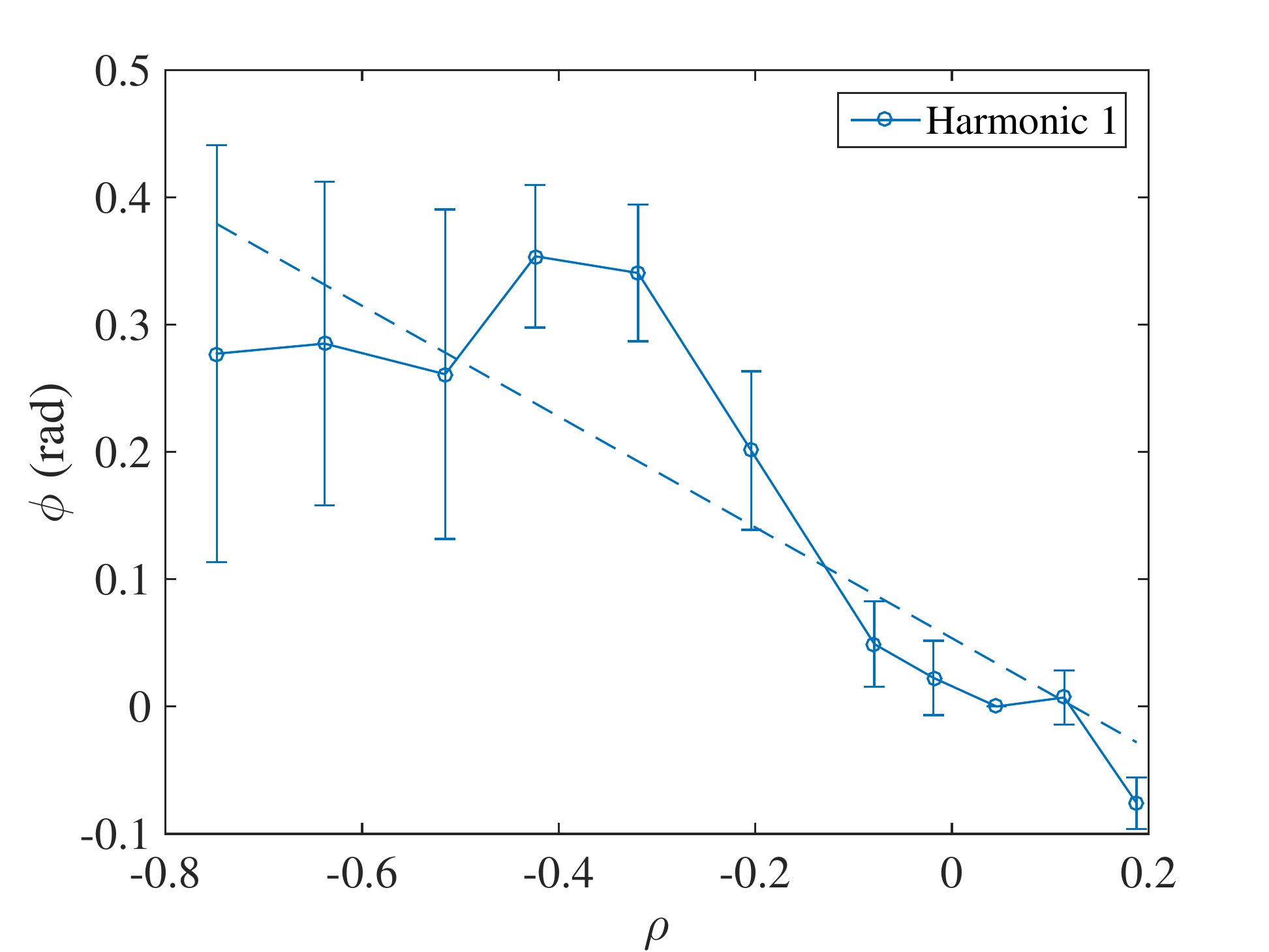}
\caption{\label{29786_pha}Cross phase, $\phi$, of the first harmonic versus radius, calculated from the $T_e(\rho,t)$ data shown in Fig.~\ref{29786_Te}.}
\end{figure}

Fig.~\ref{29786_Te} shows an example of the measured ECE time traces of $T_e(\rho,t)$.
The corresponding power spectra are shown in Fig.~\ref{29786_Te_spec}.
The frequency of the first three harmonics of the modulation frequency (360 Hz) is indicated by vertical dashed lines.
The amplitude of the first three harmonics versus $\rho$ is shown in Fig.~\ref{29786_amp}.
As the modulated ECRH power is deposited centrally, the amplitude decays from the centre of the plasma towards the edge.
The indicated error bars indicate the variation of the amplitude when the analysis is performed using successive time windows with a length of 10 ms.
The phase of the first harmonic versus $\rho$ is shown in Fig.~\ref{29786_pha}, along with a straight line fit.
For this purpose, a central ECE channel is taken as reference; evidently, for the selected reference channel the cross phase is zero.
Generally speaking, the cross phase increases from the plasma core towards the edge, indicating a delayed response to the core modulation. It is understood that the effect of the modulation propagates outward as a consequence of heat transport.
Note that the phase does not increase monotonically from the centre towards the edge; in particular, for $\rho < -0.4$ the phase seems to decrease or at least remain constant. 
This suggests that transport is not purely diffusive in this region, as pure diffusion would correspond to a monotonically increasing phase outside the power deposition region.


\begin{figure}\centering
  \includegraphics[trim=0 0 0 0,clip=,width=12cm]{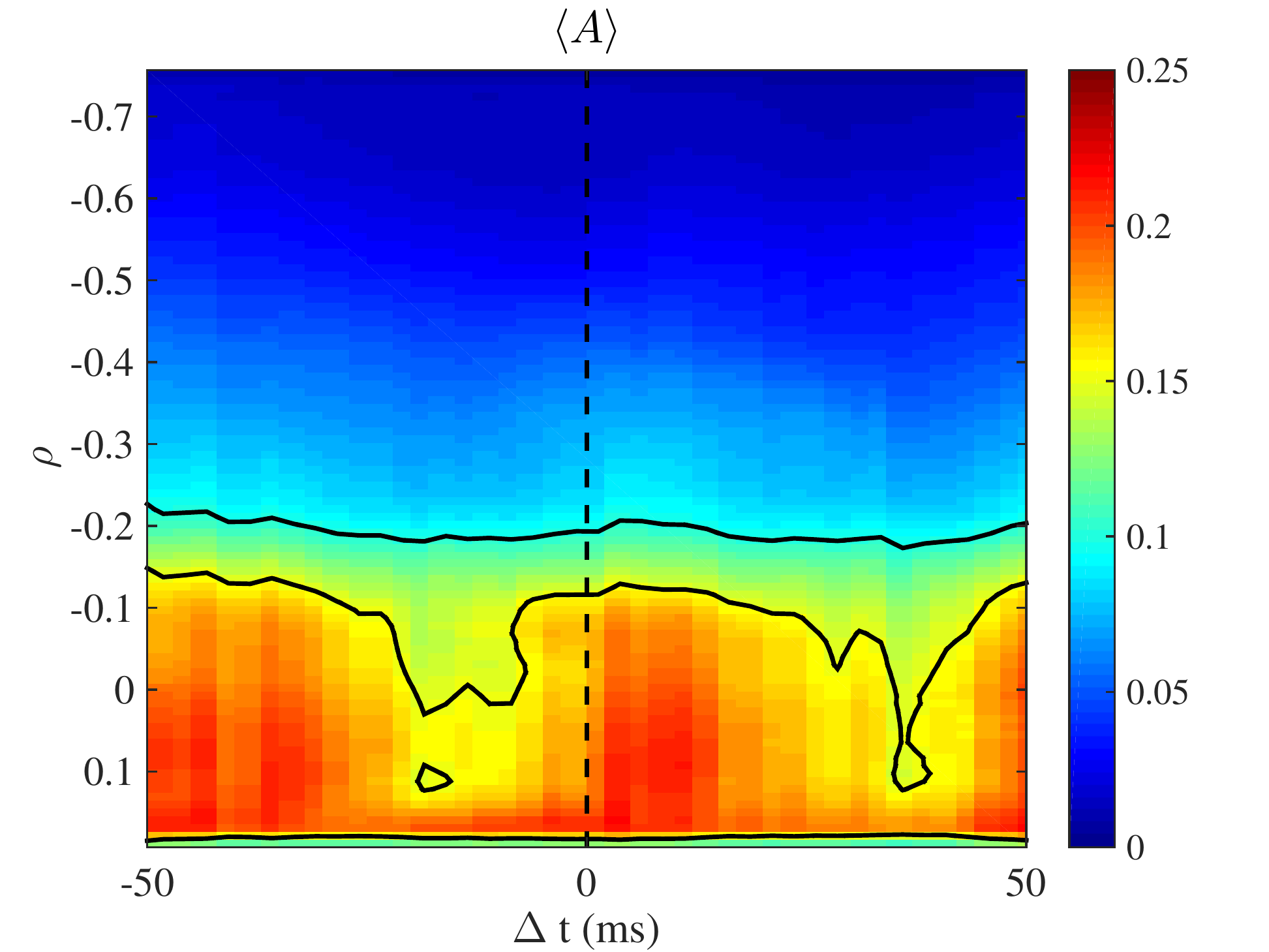}\\
  \includegraphics[trim=0 0 0 0,clip=,width=12cm]{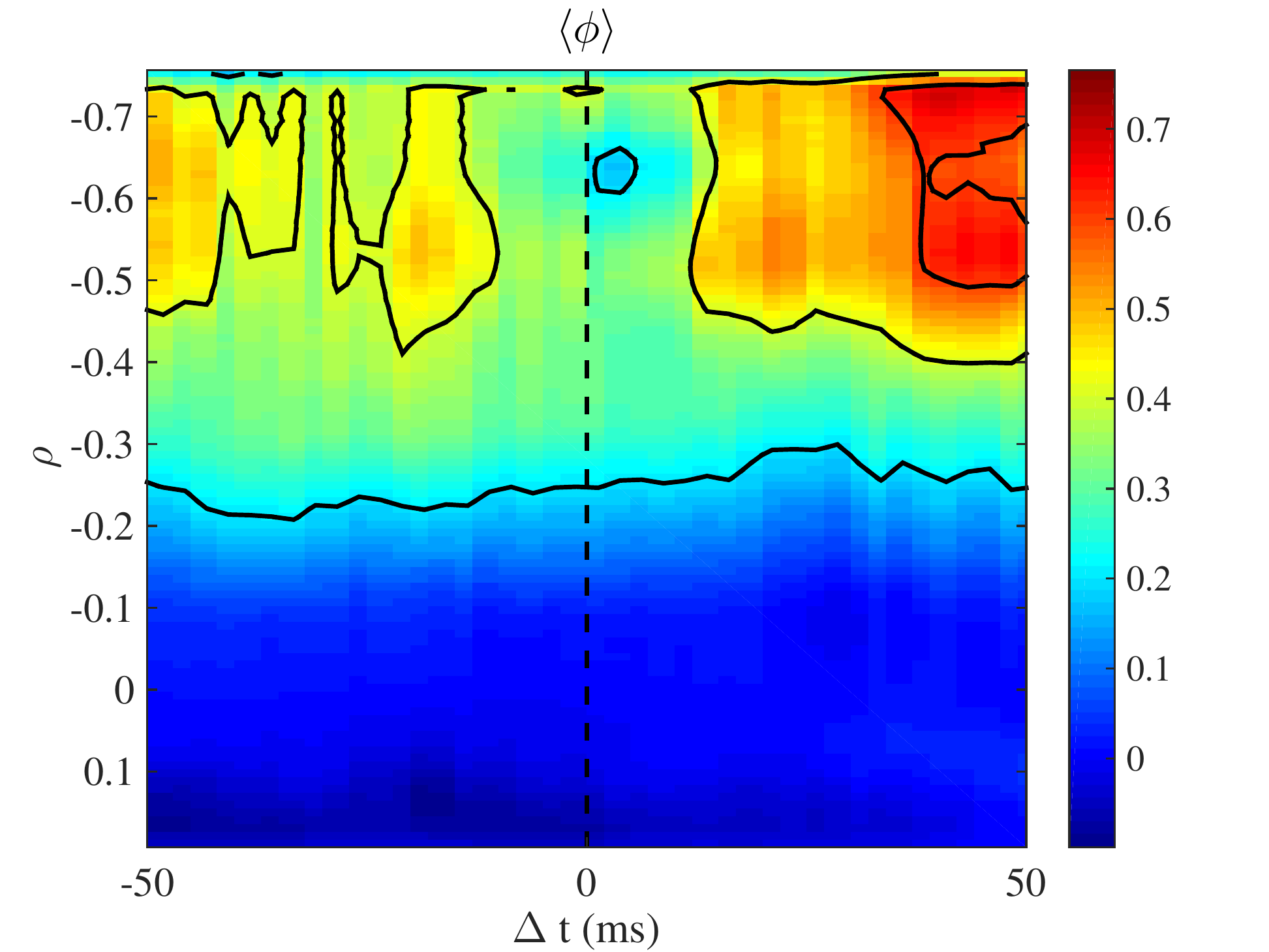}
\caption{\label{meanece}Mean ECE modulation results (top: $A$, bottom: $\phi$).}
\end{figure}

The propagation analysis is quite reproducible for the set of 8 discharges cited above. 
Hence, we use the same reference times as above and calculate discharge averaged values versus time and radius.
Fig.~\ref{meanece} shows the mean modulation amplitude and phase (relative to the central ECE channel).

Note the significant and short-lasting reduction of phase around  $\Delta t \simeq 0$ ms ($|\rho | \simeq 0.65$).
This time and place coincides rather well with the passage of the rational
$\iotabar = 3/2$, cf.~Fig.\ref{track}, and its entry into the density gradient zone.
To interpret this behavior, recall that the measured modulation amplitude and phase can be related to the effective heat diffusion coefficient via~\cite{Jacchia:1991}:
\begin{equation}\label{Jacchia}
\chi_e = \frac34 \frac{\omega_0}{\phi' (A'/A + 1/2r + n'/2n)}
\end{equation}
Thus, a drop in the radial derivative of the phase, $\phi'$, implies an increase in local $\chi$.
Fig.~\ref{meanece} shows that the phase is nearly constant over a wide area, namely $0.3 < \rho < 0.7$, for a time period lasting several tens of ms around $\Delta t = 0$. 
This observed reduction of modulation phase gradient therefore indicates a spatially and temporally localized enhancement of heat transport.
The slight temporal variation of $A$ is mainly due to a small temporal variation of $\overline {n_e}$, leading to a corresponding small variation in absorbed ECRH power.

\begin{figure}\centering
  \includegraphics[trim=0 0 0 0,clip=,width=12cm]{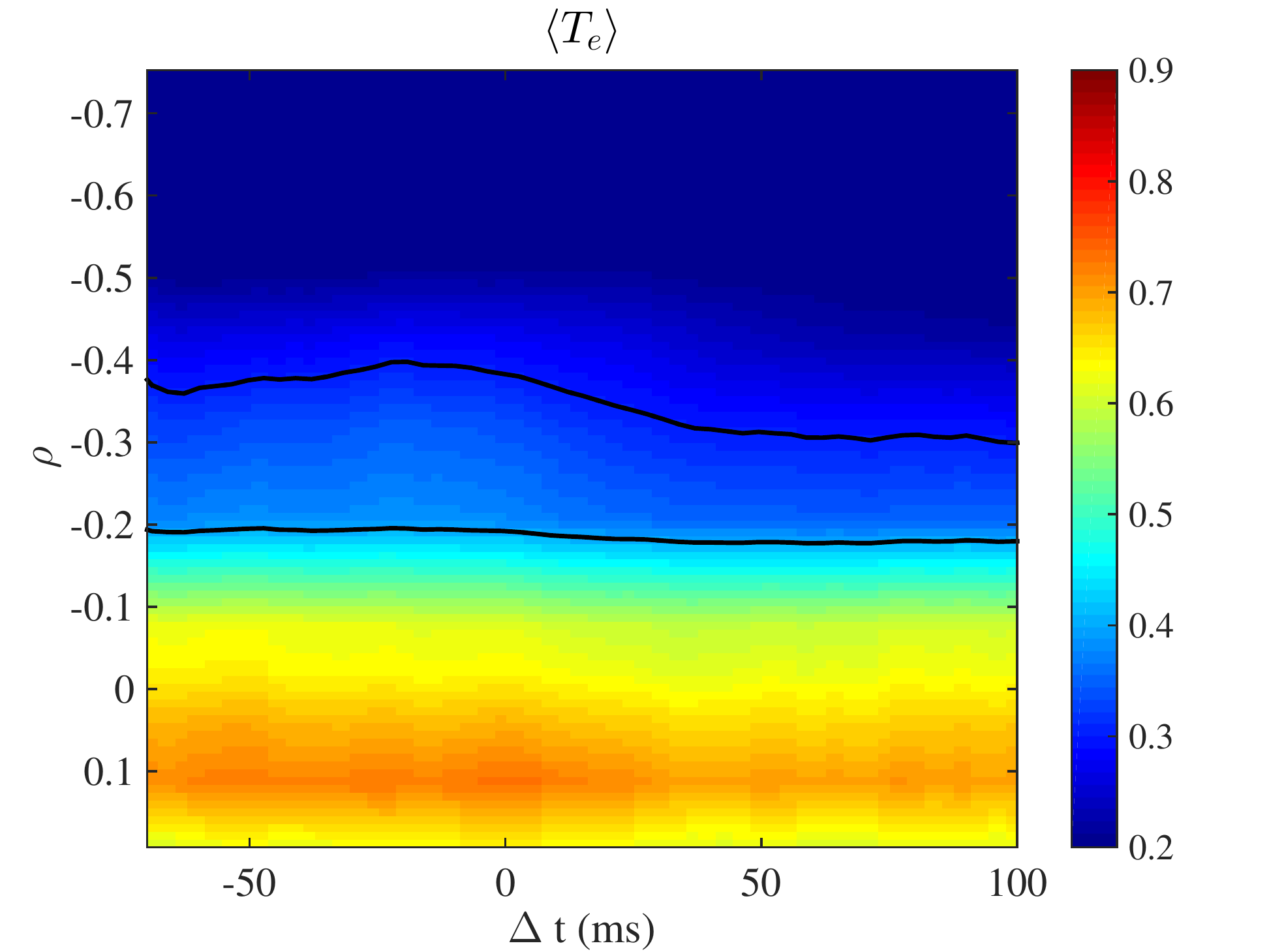}\\
  \includegraphics[trim=0 0 0 0,clip=,width=12cm]{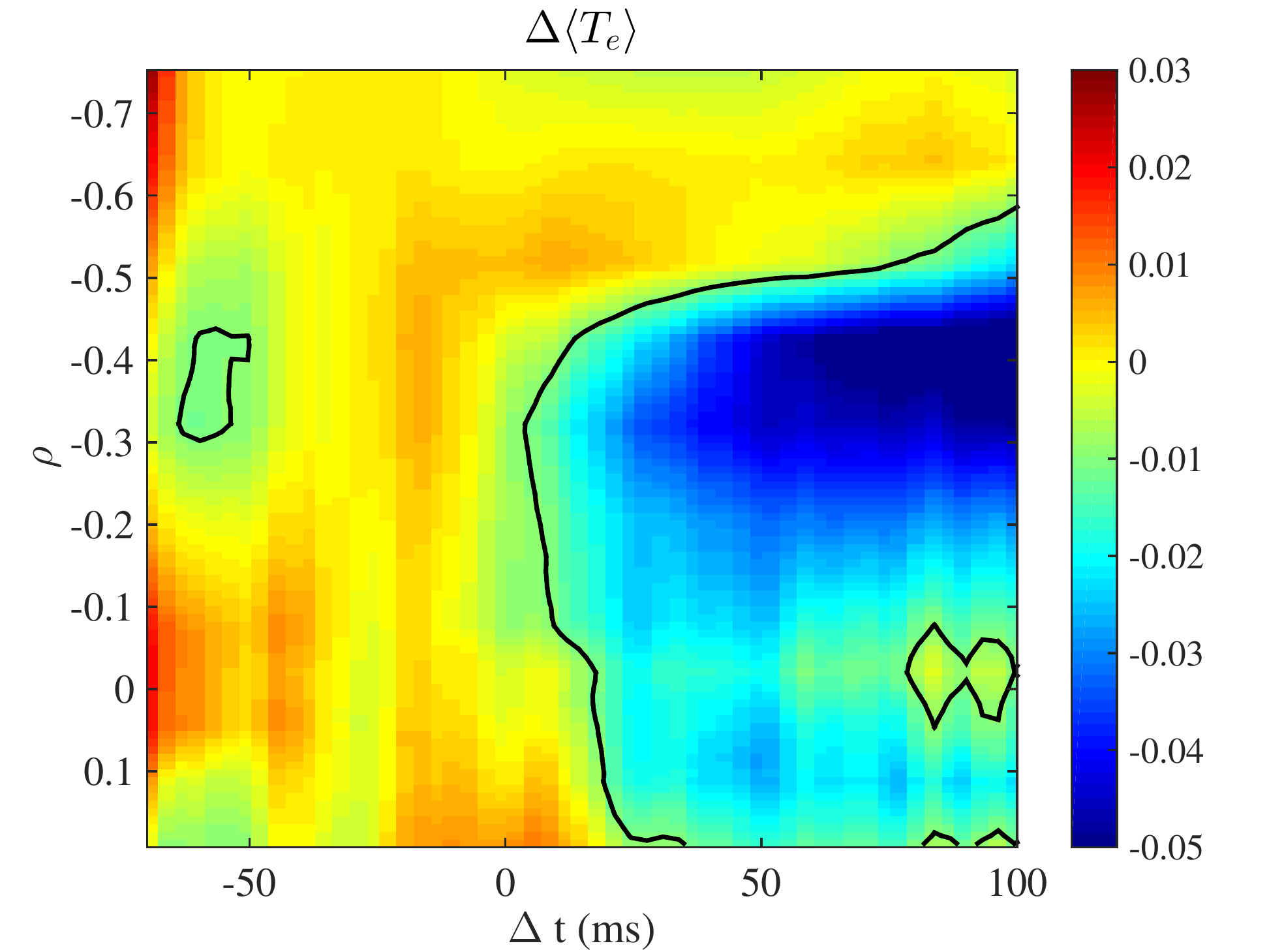}
\caption{\label{mean_Te}$\langle T_e(\rho,t) \rangle$, averaged over 8 similar discharges, and $\Delta \langle T_e(\rho,t) \rangle$. Units: keV.}
\end{figure}

Fig.~\ref{mean_Te} shows the time evolution of the measured electron temperature, $\langle T_e(\rho,t) \rangle$, averaged over the 8 mentioned discharges, and its temporal variation, $\Delta \langle T_e(\rho,t) \rangle = \langle T_e(\rho,t) \rangle - \overline {\langle T_e(\rho,t) \rangle}$, where the overline indicates a time average taken over the time window $-50 \le \Delta t \le 0$ ms.
By definition, this quantity is near zero for $-50 < \Delta t < 0$ ms, but a significant temperature drop ($|\Delta T_e|  \le 50$ eV) is visible for $\Delta t > 0$, starting at around $\rho = 0.35$ and propagating both inward and outward, suggesting a release of heat from the core towards the edge.

\clearpage
\subsection{ECE modulation amplitude variations}

\begin{figure}\centering
  \includegraphics[trim=0 0 0 0,clip=,width=12cm]{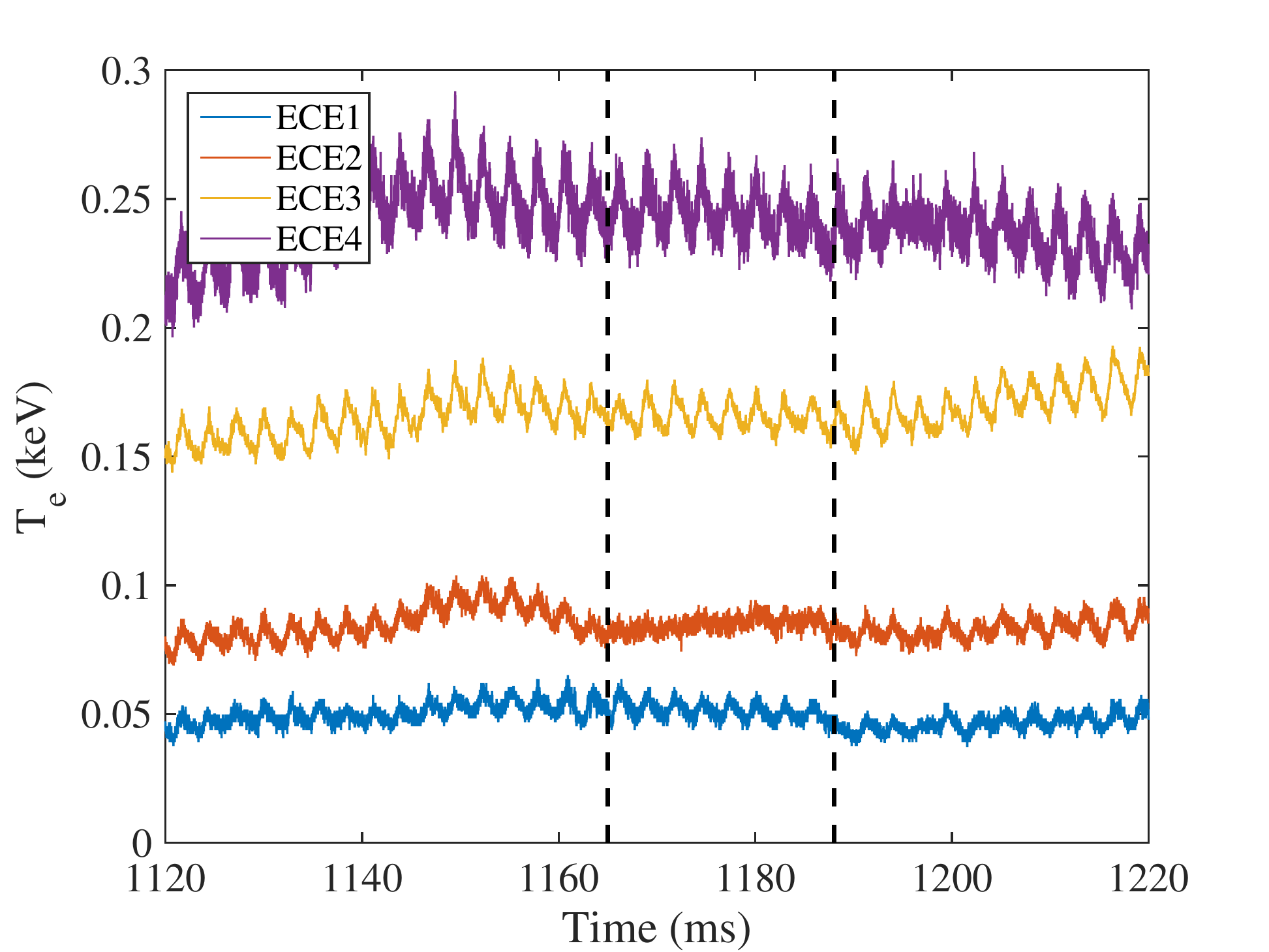}
\caption{\label{29780_ECE}Measured ECE temperature (discharge 29780). Shown are ECE channels ECE1 through ECE4, at 
$\rho = -0.75,-0.64,-0.53,-0.44$, respectively. Units: keV.}
\end{figure}

Fig.~\ref{29780_ECE} shows the signals of the outermost 4 ECE channels (top).
The modulation signal `disappears' in ECE channel 2, for $1165 < t < 1188$ ms, the time interval indicated by the vertical dashed lines.
The length of this time interval coincides with the half period of the velocity variation observed by DR (not shown for this discharge, but similar in shape to Fig.~\ref{COG_Hurst}, discharge 29782), which in turn is related to the radial extension of the magnetic island~\cite{Waelbroeck:2009,Nishimura:2010}.
Thus, a possible explanation for this phenomenon is that the modulated component of $T_e$ is weak inside the island O-point region as it sweeps past the observation point.
Even so, the modulation of the centrally deposited ECRH power does reach adjacent ECE channels, so this effect occurs only at specific radii. No important profile flattening is observed, within the resolution of the diagnostic~\cite{Hill:2015}.
Thus, the reason for the disappearance of the modulated component could be related to the magnetic topology, the island being isolated from the main plasma, and/or the relatively strong sheared flow associated with the rational, in combination with a significant local gradient.

\begin{figure}\centering
  ~\includegraphics[trim=0 0 0 0,clip=,width=12cm]{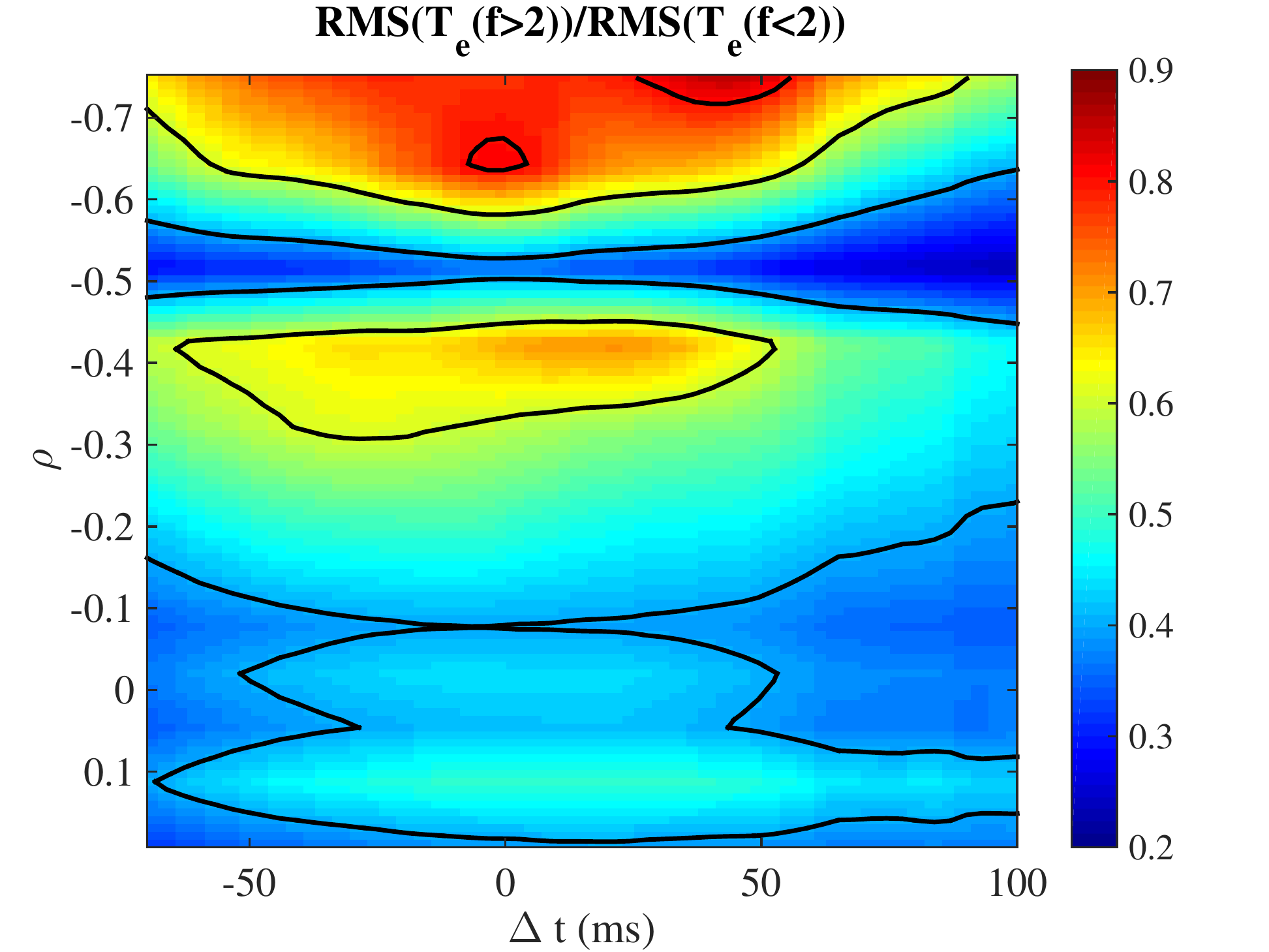}
\caption{\label{ECE_noise_vs_amp}Ratio $R$ of high frequency noise versus modulation amplitude, measured by the ECE system (see text), plotted versus time and radius (averaged over 8 discharges).}
\end{figure}

Again, to show how systematic this reduction of the modulated component of $T_e$ is, we calculate an average over the 8 mentioned discharges.
For this purpose, Fig.~\ref{ECE_noise_vs_amp} displays the mean ratio $R$, a quantifier of the degree of isolation from the temperature modulation, versus time and radius. This ratio $R = R_1/R_2$ is defined as follows. 
$R_1$ is defined as the root mean square (RMS) amplitude of the measured $T_e(\rho,t)$ data, after high-pass filtering with a cutoff frequency of 2 kHz. 
$R_2$ is defined as the RMS amplitude of the measured $T_e(\rho,t)$ data, after band-pass filtering in the frequency range $0.1 < f < 2$ kHz.
Thus, $R_1$ is a measure of the noise amplitude while $R_2$ is a measure of the modulation amplitude (the modulation frequency being 0.36 kHz).

As argued above, the island O-point region is topologically separate from the main plasma and/or characterized by a significant sheared flow (at least, in the present case), which would reduce the modulated component of $T_e$ in this region.
Thus, the regions in Fig.~\ref{ECE_noise_vs_amp} with high values of $R$ (i.e., low modulation amplitude) would correspond to island O-point regions and/or sheared flow regions.
Such regions with high $R$ can of course only be observed clearly (at the fixed ECE measurement positions) if the island is large and stationary, and/or the sheared flow is strong.
Apparently, this condition is met when the rational 3/2 surface (moving outward) reaches $|\rho| > 0.65$, which is the density gradient region.
At earlier times, when the rational is further inward, the island may be smaller and/or rotating, such that $R$ is smaller. 
The zone of relatively high $R$ seen at $|\rho| \simeq 0.4$ is possibly associated with another rational ($\iotabar = 7/5$); cf.~Fig.~\ref{track}.
In view of the above argument, this 7/5 mode must also be stationary to be visible; yet it is not located in the gradient region. 
Therefore, we speculate that its apparent stationarity is a consequence of mode locking with the stationary 3/2 mode.
Note that the total magnetic field, $|B|$, varies by less than 0.5\% due to the varying plasma current, so that the radial locations of the ECE channels, inversely proportional to $|B|$, may be considered fixed.

\clearpage
\section{Modeling}\label{modeling}

Topological analysis of the flow structures of plasma turbulence~\cite{Carreras:2014} has shown 
that typically, low order rational surfaces are associated with small transport barriers. 
These are created by zonal flows, induced by flow eddies associated with the rational surfaces. 
These small transport barriers tend to trap particles, leading to local subdiffusive behavior, as reported in \cite{Newman:1996}. 
Therefore, a reduction of the Hurst coefficient below 0.5 is expected in the barrier regions.
On the other hand, experimental studies have shown that $H$ is systematically large in the plasma edge region~\cite{Carreras:1998b} and varies with the global gradients~\cite{Carralero:2011}.

To study the correlation between the zonal flows associated with low order rationals and the Hurst coefficient, we have performed a simulation of the magnetohydrodynamic turbulence induced by resistive interchange modes in a periodic cylinder. 
Details of this type of modeling can be found in Ref.~\cite{Garcia:2001}. 
We have not attempted to match the experimental situation closely, due to the difficulty of reproducing a gradual modification of the rotational transform in time, induced by an increasing plasma current. 
Here, we limit ourselves to studying the possible relation between low order rational surfaces, zonal flow generation and the Hurst coefficient.
For this purpose, we have taken one of the standard rotational transform profiles of TJ-II, the case labelled 100\_46 in Fig.~\ref{iota}, using a pressure profile that is close to the experimental one. 
Because the rotational transform has low shear, the low order rational surfaces produce magnetic islands of a significant width, facilitating the detection of regions with potential subdiffusion. 
 
The main parameters used in these calculations are: $\beta_0 = 0.001$, inverse aspect ratio $\epsilon=a/R_0 = 0.15$ and Lundquist number $S = \tau_R/\tau_A = 10^5$. 
For these parameters, this configuration is unstable to resistive interchange modes. 
Here, $\beta_0$ is the ratio of the plasma pressure and the magnetic pressure at the plasma axis, $\beta_0=2\mu_0p(0)/B_\zeta^2$, $\tau_R$ is the resistive time at the magnetic axis, $\tau_R = \mu_0a^2/\eta(0)$, where $\eta(0)$ is the resistivity at the magnetic axis and $\tau_A$ is the Alfven time, $\tau_A = R_0 \sqrt{\mu_0 m_in_i}/B_\zeta$.

\begin{figure}\centering
  \includegraphics[trim=0 0 0 0,clip=,width=16cm]{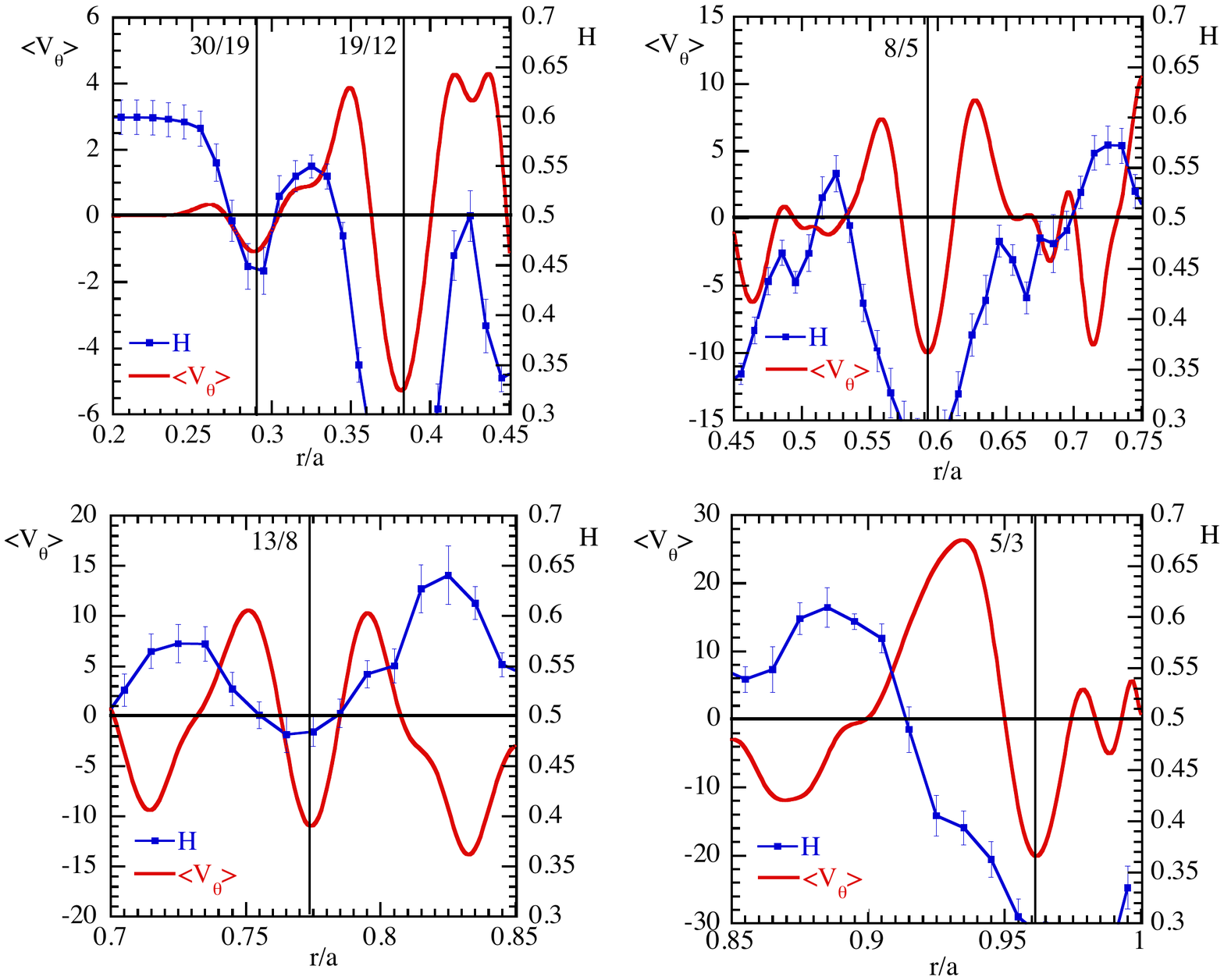}
\caption{\label{Ben_a}Modeling results: mean poloidal flow and Hurst coefficient near various rationals.}
\end{figure}

We have calculated the Hurst coefficient at 100 equally spaced radial locations over a period of time of $4\tau_R$, long enough to obtain asymptotic behavior. We also calculated the angular and time averaged poloidal flow. 
The Hurst coefficient and the averaged poloidal flow are shown in Fig.~\ref{Ben_a}. 

At each of the main rational surfaces, we can see that the averaged poloidal flow has a minimum, while the flow has strong shear in the region of the magnetic island associated with each resonant surface. It is also clear that the $H$ coefficient has a minimum in the same region, with values below 0.5. The $H$ coefficient increases above 0.5 as we move out of the region of the resonant surface. This structure is fairly systematic at all the low resonant surfaces present in the plasma and they are consistent with the measurements shown in Fig.~\ref{COG_Hurst}.
The experiment also shows a clear enhancement of $H$ inside from the rational surface, cf.~Fig.~\ref{mean_COG_H}.

When comparing with the experiment, on should remember that transport consists of both turbulent and collisional transport, the latter being induced by collisions~\cite{Sanchez:2001}. 
Collisional transport has a diffusive nature and tends to dominate subdiffusive processes, making it difficult, in general, to observe significant regions with $H < 0.5$ in the experiment. 
Nevertheless, times with low values of $H$ are visible in Fig.~\ref{COG_Hurst}, and $H$ is systematically moderate (of the order of 0.5 or slightly below) in a zone associated with the $\iotabar = 3/2$ island, cf.~Fig.~\ref{mean_COG_H}.

\begin{figure}\centering
  \includegraphics[trim=0 0 0 0,clip=,width=12cm]{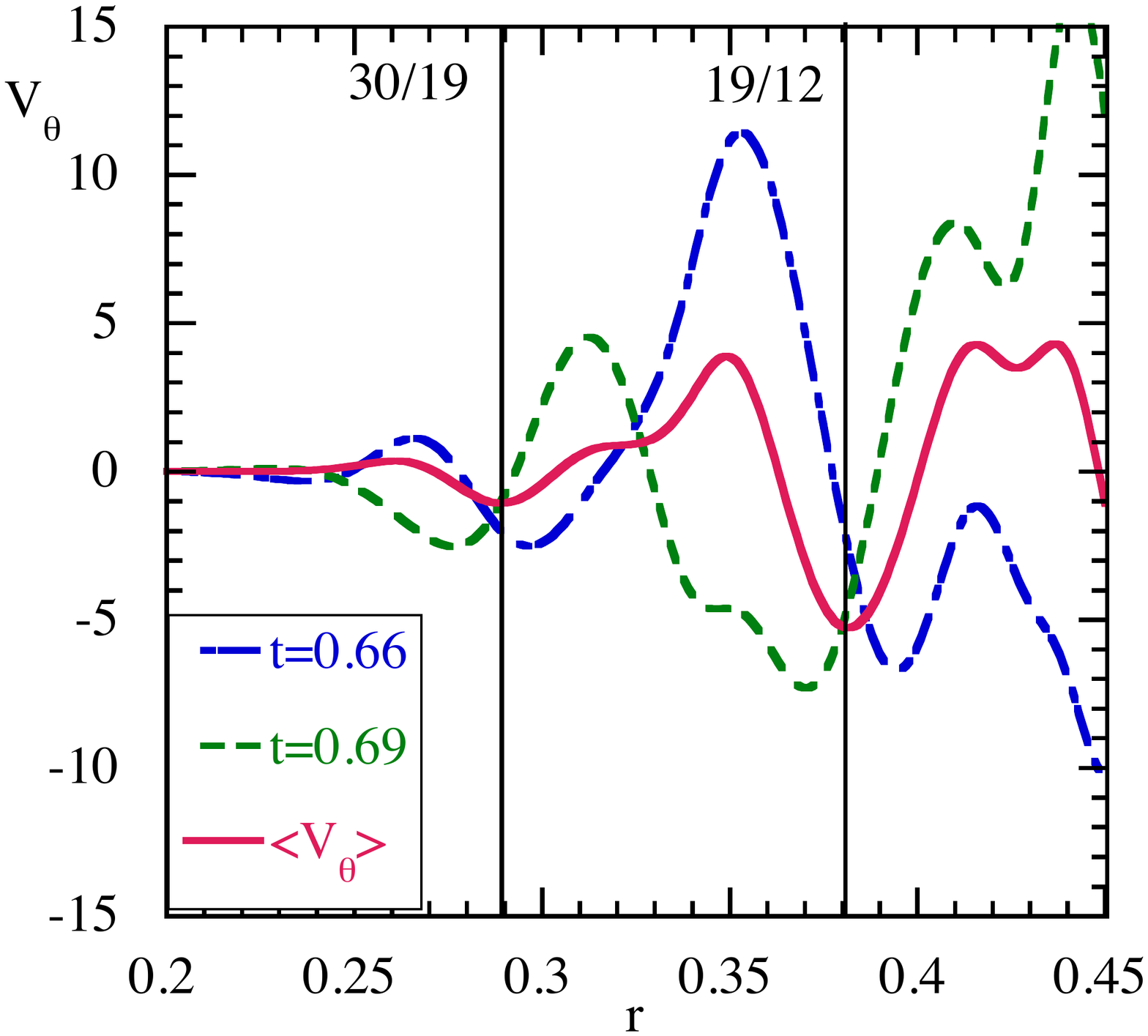}
\caption{\label{Ben_b}Modeling results: instantaneous poloidal flows at two time points, and mean poloidal flow.}
\end{figure}

The averaged flow shown in Fig.~\ref{Ben_a}, $\langle V_\theta \rangle$, is calculated by averaging the poloidal flow over the poloidal and toroidal angles and over a period of time such that the dynamical calculation and the corresponding zonal flow are in steady state. 
Note that this mean flow is different from the perpendicular flow measured in the experiment by DR, which is instantaneous and local, and hence may be very sensitive to the precise poloidal and toroidal location of the measurement (i.e., the phase of the stationary island).
If we compare the averaged flow near a singular surface to the instantaneous poloidal flow, cf.~Fig.~\ref{Ben_b}, one observes that the latter is not symmetric with respect to the singular surface, and that the asymmetry may vary.
This would explain the asymmetric appearance of the poloidal flow with respect to the rational surface observed in the experiment.

\begin{figure}\centering
  \includegraphics[trim=0 0 0 0,clip=,width=12cm]{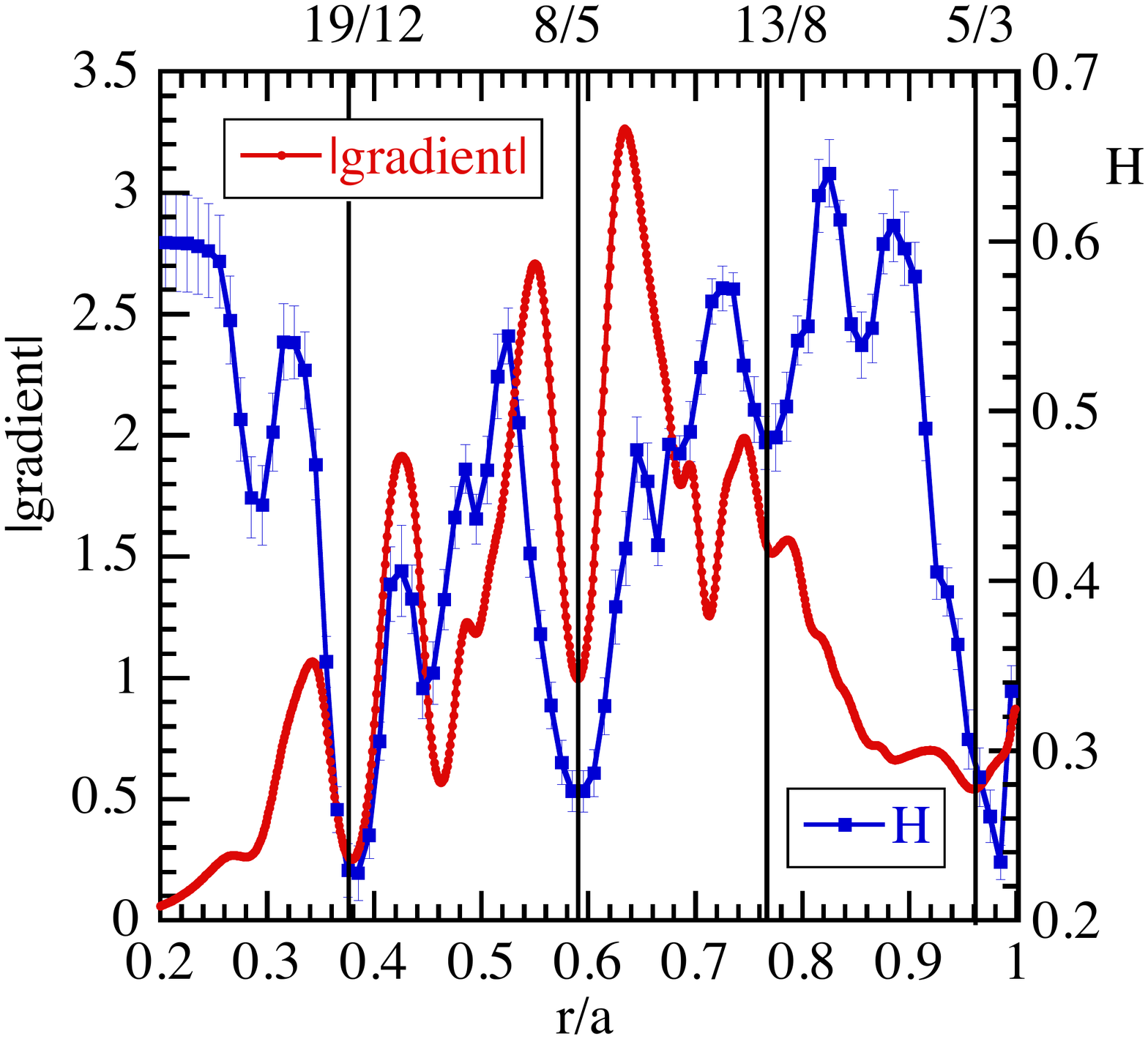}
\caption{\label{Ben_c}Modeling results: absolute value of the gradient and Hurst coefficient.}
\end{figure}

The $H$ coefficient is determined by at least two effects.
One is the generation of long range correlations by avalanches, and the other is their suppression due to decorrelation by sheared flow. Avalanches are non-local phenomena, but in the plasma core region avalanches are relatively rare and they tend to correlate with the gradient, as they originate there.
The edge, however, is crossed by many avalanches initiated in the entire plasma, so no relation is expected between the local gradient and the $H$ exponent. 
At the plasma edge, the $H$ exponent tends to be above $0.5$~\cite{Carreras:1998b}. 
In Fig.~\ref{Ben_c} we show the gradient of the pressure, averaged over the flux surface and over time, 
as a function of the radius, together with the $H$ coefficient, to illustrate this situation.
At the singular surface, the density gradient is low, and due to the presence of the islands and the associated average sheared flow, the $H$ coefficient is reduced below $0.5$.
Immediately outside the islands, where the sheared flow goes to zero, the density gradient is high, correlated with an increase of $H$ above $0.5$, at least in the plasma core region. 
In the plasma edge region, $H$ tends to rise above $0.5$ globally, while it is reduced below $0.5$ only at the singular surfaces, mainly $\iotabar = 5/3$ in this case, without any clear correlation with the gradient.

\clearpage
\section{Discussion}\label{discussion}

Doppler reflectometry (DR) has detected a structure in $v_\perp$ (equivalent to $E_r$) that is systematically associated with the $\iotabar = 3/2$ rational surface, which is purposely and slowly scanned outward in the experiments discussed here.
We take the time of maximum temporal variation of $v_\perp$ as a reference time ($\Delta t = 0$); this time is closely associated with, though not identical to, the time of passage of the rational surface itself~\cite{Nishimura:2010}.
When the rational surface crosses the DR location of channel 1, the Hurst coefficient, $H$, is reduced (cf. Fig.~\ref{COG_Hurst}).
This is consistent with the systematic reduction of $H$ at rational surfaces observed in the model discussed in Section~\ref{modeling}.
A reduction of $H$ is associated with subdiffusive transport, implying that possibly, the island constitutes a (minor) transport barrier~\cite{Beyer:2000,LopezBruna:2009}, which in turn may be related to the relatively good heat confinement observed in Fig.~\ref{mean_Te} for $\Delta t < 0$.

If the location of the singular surface is associated with the minimum of the Hurst coefficient, Fig.~\ref{COG_Hurst} indicate that the passage of the singular surface occurs slightly before (i.e., it's outward) from the point of maximum slope of $v_\perp$.

Some MHD mode oscillations ($\sim 30$ kHz) are observed (bolometry, DR) for $\Delta t < 0$, which are reduced or disappear afterwards.
As these mode oscillations gradually reduce their mean frequency, it is clear that they are associated with rational surfaces that are moving outward to areas of the plasma with smaller mean rotation velocity~\cite{Milligen:2012}.
The proposed explanation for the disappearance is that the $\iotabar = 3/2$ island is small for $\Delta t < 0$, but large and stationary for $\Delta t > 0$.
This is consistent with the observation $v_\perp \simeq 0$ at the rational surface in the density gradient region, cf.~Fig.~\ref{COG_Hurst}.

This interpretation, which relies on the existence of a large island in the gradient region, finds support from the observation of specific radial zones where the ECRH modulation penetrates only weakly (Fig.~\ref{ECE_noise_vs_amp}), which are interpreted as topologically isolated island O-point regions~\cite{Spakman:2008} and/or zones with considerable velocity shear.

For $0 < \Delta t \le 25$ ms, the Hurst coefficient is large ($H(\rho \simeq 0.75) \simeq 0.65-0.7$), although somewhat further inward, $H$ stays moderate ($H(\rho \simeq 0.68) \simeq 0.5$).
A representation of the evolution of the mean ECE temperature $\langle T_e \rangle$ and its variation $\Delta \langle T_e \rangle$ (Fig.~\ref{mean_Te}) suggests heat loss occurring for $\Delta t> 0$, starting in the core and expanding outward, as visualized in the plot of  $\Delta \langle T_e \rangle$ as a dark blue area.
Simultaneously, the $T_e$ modulation phase delay gradient ($d \phi / dr$) is reduced in the region $0.3 < \rho < 0.7$. 
We note that the (incremental) heat diffusivity is proportional to the inverse of the gradient of the phase~\cite{Mantica:2006b}, 
so that the decrease of $d\phi/dr$ suggests enhanced heat transport and stiffness in this region (inward from the island).
This observation establishes a clear link between (memory effects associated with) density fluctuations and heat transport.

\clearpage
\section{Conclusions}\label{conclusions}

In these experiments at TJ-II, an externally driven OH current was used to modify the rotational transform profile, $\iotabar(\rho)$, and slowly move the rational surfaces outward through the plasma.
As the main rational $\iotabar = 3/2$ moved past the DR observation points, the perpendicular propagation velocity of the turbulence, $v_\perp$ -- equivalent to the radial electric field, $E_r$ -- was found to experience a significant `wiggle'.
Taking the time of strongest temporal variation of $v_\perp$ as a reference, 8 similar discharges were analyzed.

Small rotating MHD modes, were detected by bolometry, with decreasing frequency, consistent with the radial outward motion of the rational surface locations.
Once the $\iotabar = 3/2$  rational was well inside the density gradient region, the corresponding island grew and was found to be stationary.
No clear flattening of profiles was observed, which could be consistent with expectations for small magnetic shear~\cite{Waelbroeck:2009}.
However, it was found that the modulated heat deposited in the core plasma only penetrated weakly in the zone associated with the stationary island.

The Hurst coefficient was found to be low ($H < 0.5$) at the rational surface (Fig.~\ref{COG_Hurst}) and
systematically high ($H \simeq 0.65-0.7$) inward from the $\iotabar = 3/2$ rational surface (Fig.~\ref{mean_COG_H}), as long as the rational surface was located well inside the density gradient region. 
The analysis of ECE data revealed that radial heat transport was enhanced in the region inward from the large island.

Simulations of magnetohydrodynamic turbulence induced by resistive interchange modes reproduced some of these observations qualitatively, namely: (1) the variation of poloidal flow profiles associated with rational surfaces, and (2) the variation of the Hurst coefficient associated with rational surfaces and local pressure gradients.

Thus, this work establishes a clear correlation between density fluctuation long range temporal correlation properties (quantified via the Hurst coefficient) and enhanced heat transport in the presence of a large island, possibly providing the first direct observation of the turbulent mechanism behind anomalous heat transport.

\section*{Acknowledgements}
Research sponsored in part by the Ministerio de Econom\'ia y Competitividad of Spain under project Nrs.~ENE2012-38620-C02-01, ENE2012-38620-C02-02, ENE2012-30832, and ENE2013-48109-P.
This work has been carried out within the framework of the EUROfusion Consortium and has received funding from the Euratom research and training programme 2014-2018 under grant agreement No 633053. 
The views and opinions expressed herein do not necessarily reflect those of the European Commission.

\clearpage
\section*{References}


\end{document}